%% file: main.tex
\documentclass[acmsmall,screen,nonacm]{acmart}

\AtBeginDocument{%
  }

\usepackage[utf8]{inputenc}
\usepackage{amsmath}
\usepackage{amsfonts}
\usepackage{amsthm}
\usepackage{xcolor}
\usepackage{cleveref}
\usepackage[english]{babel}
\usepackage{xspace}
\usepackage[T1]{fontenc}
\usepackage{refs}
\usepackage{pifont}
\usepackage{multirow}
\usepackage{subcaption}
\usepackage{mathpartir}
\usepackage{soul}
\usepackage{rotating}
\usepackage{wrapfig}

\usepackage[leftbars,color]{changebar}

\cbcolor{red}

\usepackage{tikz}
\usepackage{tikz-3dplot}
\usetikzlibrary{shapes.geometric,shapes.misc,fit,positioning,backgrounds,intersections,calc}
\pgfdeclarelayer{background layer}
\pgfdeclarelayer{foreground layer}
\pgfsetlayers{background layer,main,foreground layer}
\makeatletter
\newcommand\currentcoordinate{\the\tikz@lastxsaved,\the\tikz@lastysaved}

\usepackage[noline,noend,ruled,linesnumbered]{algorithm2e}
\definecolor{light-gray}{gray}{0.9}
\SetInd {0.0em}{0.7em} % indentation for algorithm.

\usepackage{semantic}
\usepackage{url}

\usepackage{listings}
\lstdefinestyle{base}{
  language=C,
  emptylines=1,
  breaklines=true,
  basicstyle=\color{black}\fontfamily{pzc}\selectfont\tt,
  commentstyle=\color{gray}\rm\itshape,
  keywordstyle=\rm\bfseries,
  identifierstyle=\rm\itshape,
  escapechar=@,
  morekeywords=[1]{then,do,halt,class,function,unint},
  morekeywords=[2]{assert},
  morekeywords=[3]{requires,ensures},
  keywordstyle	= [2]\color{red}\bf,
  keywordstyle	= [3]\bf\color{gray},
  numberstyle=\footnotesize\color{gray},
  moredelim=**[is][\bf\color{green}]{@!}{!@},
  moredelim=**[is][\bf\color{red}]{@?}{?@},
  literate=
    {<=}{{$\leq$}}1
    {>=}{{$\geq$}}1
    {!}{{$\neg$}}1
    {!=}{{$\neq$}}1
    {||}{{$\lor$}}1
    {&&}{{$\land$}}1
    {->}{{$\rightarrow$}}1
    {_1}{$_{1}$}2
    {_2}{$_{2}$}2
    {_3}{$_{3}$}2
}
\newcommand{\cinline}[1]{\mbox{\lstinline[style=base,mathescape]{#1}}}

\newcommand{\Omit}[1]{}

\newcommand{\LT}{\mathrm{LT}}
\newcommand{\defeq}{\triangleq}
\newcommand{\floor}[1]{\left\lfloor #1 \right\rfloor}

\newcommand{\lproj}[3]{\mathrm{LProj}(#1, #2, #3)}
\newcommand{\proj}[2]{\mathrm{Proj}(#1, #2)}
\DeclareMathOperator*{\argmax}{arg\,max}
\newcommand{\LM}{\mathrm{LM}}
\newcommand{\MOLE}{\ll}
\newcommand{\MOGE}{\gg}

\newcommand{\effdegmw}[3]{\text{eff.deg}_{#1}^{#2}(#3)}

\newcommand{\Tool}{{\textsc{AutoBound}}\xspace}

\newif\iflong
\longtrue
\iflong
\newcommand{\refappendix}[1]{\sectref{#1}}
\usepackage[appendix=append]{apxproof} %This will keep the proofs
\else
\newcommand{\refappendix}[1]{the supplementary materials}
\usepackage[appendix=strip]{apxproof} %This will omit the proofs
\fi

\newtheoremrep{theorem}{Theorem}[section]
\newtheoremrep{corollary}{Corollary}[theorem]
\newtheorem{definition}[theorem]{Definition}
\newtheoremrep{lemma}[theorem]{Lemma}
\newtheorem{example}[theorem]{Example}

\newtheorem*{remark}{Remark}

\title{Optimal Symbolic Bound Synthesis}

\author{John Cyphert}
\affiliation{
  \institution{University of Wisconsin--Madison}            %% \institution is required
   \country{United States of America}                    %% \country is recommended
}
\author{Yotam Feldman}
\affiliation{
  \institution{Tel Aviv University}            %% \institution is required
   \country{Israel}                    %% \country is recommended
}
\author{Zachary Kincaid}
\affiliation{
  \institution{Princeton University}            %% \institution is required
   \country{United States of America}                    %% \country is recommended
}
\author{Thomas Reps}
\affiliation{
  \institution{University of Wisconsin--Madison}            %% \institution is required
  \country{United States of America}                    %% \country is recommended
}
\date{}

\begin{document}

\input{abstract}

\maketitle

\input{intro}
\input{overview}
\input{background}
\input{reduction}
\input{saturation}
\input{effective_degree}
\input{experiments}
\input{relatedwork}
%\input{conclusion}

\bibliographystyle{ACM-Reference-Format}
\bibliography{references}

\end{document}

%% file: abstract.tex
\begin{abstract}
The problem of finding a \emph{constant} bound on a term given a set of assumptions has wide applications in optimization as well as program analysis. However, in many contexts the objective term may be unbounded. Still, some sort of \emph{symbolic} bound may be useful. In this paper we introduce the \emph{optimal symbolic-bound synthesis problem}, and a technique that tackles this problem for \emph{non-linear arithmetic} with function symbols. This allows us to automatically produce symbolic bounds on complex arithmetic expressions from a set of both equality and inequality assumptions. 
Our solution employs a novel combination of powerful mathematical objects---\emph{Gr\"obner bases} together with \emph{polyhedral cones}---to represent an infinite set of implied inequalities. We obtain a sound symbolic bound by \emph{reducing} the objective term by this infinite set. 

We implemented our method in a tool, \Tool, which we tested on problems originating from real Solidity programs. We find that \Tool yields relevant bounds in each case, matching or nearly-matching upper bounds produced by a human analyst on the same set of programs.

% The problem of finding a \emph{constant} bound on a term given a set of assumptions has wide applications in optimization as well as program analysis. However, in many contexts the objective term may be unbounded, still some sort of \emph{symbolic} bound may be useful. In this paper we introduce and partially address the optimal symbolic bound synthesis problem (OSB). Moreover, our method considers the case when the set of assumptions as well as the objective term are \emph{non-linear}. Our solution employs the combination of the powerful mathematical objects Gr\"obner bases and polyhedral cones to represent an infinite set of implied inequalities. We obtain a sound symbolic bound by \emph{reducing} the objective term by this infinite set. We implemented our method in a tool \Tool, which we tested by translating real Solidity programs to safe assumptions and asking \Tool to find safe upper bounds on terms over program variables. We find that \Tool yields relevant bounds in each case, matching or nearly matching upper bounds produced by a human analyst on the same set of programs.
\end{abstract}

%% file: intro.tex
\section{Introduction}\label{Se:introduction}
In this paper we introduce and address the following problem, which we call the \emph{optimal symbolic-bound synthesis (OSB)} problem:
\begin{quote}
    \textbf{Given} a (potentially non-linear) formula $\phi$ representing assumptions and axioms and an objective term $t$, \textbf{find} a term $t^{*}$ such that
    \begin{enumerate}
    \item (Bound) $\phi \models t \le t^{*}$
    \item (Optimality) For every term $s$ that satisfies the first condition, $t^{*} \preceq s$ holds,
      where $\preceq$ represents some notion of ``term desirability.''
    \end{enumerate}
\end{quote}
A solution to this problem has many applications in the automatic analysis of programs.
For example, a common program-analysis strategy is to extract the semantics of a program as a logical formula, say $\phi$. However, such a formula can contain temporary variables and disjunctions, and is therefore it is difficult for a human to understand the dynamics of the program from $\phi$. An instance of the OSB problem allows a user to specify a term of interest $t$, e.g. representing some resource, such as time, space or the value of some financial asset, 
and a term order $\preceq$ 
that strictly favors terms only over input parameters. In this instance an OSB solver
produces a sound upper-bound $t^{*}$ 
on the resource $t$ with all the temporary variables projected out.

The problem of finding \emph{constant} bounds on a term $t$ given assumptions $\phi$ is commonly addressed in the field of optimization.
However, in the context of program analysis we are often not interested in constant bounds on such a term---either because $t$ is unbounded, or because the bound on $t$ is so loose as to be uninformative.
An alternative approach---the one adopted in this paper---is to find a \emph{symbolic} bound, given assumptions $\phi$.

The OSB problem as given above is very general. Namely, we have yet to specify any restrictions on $\phi$, $\models$, or the term-desirability order $\preceq$. In future we would like others to consider methods to address OSB problems for various instantiations of $\models$ and $\preceq$.
In this paper, we consider the OSB problem in the context in which $\phi$, $\models$ and $t$ are interpreted over \emph{non-linear} arithmetic, and $\preceq$ gives a intuitive, human notion of a simpler term. Moreover, we do not place a restriction on the form of $\phi$. That is, $\phi$ is an arithmetic formula with the usual boolean connectives.

This setting introduces the significant challenge of non-linear arithmetic reasoning: 
(i) in the case of rational arithmetic, it is undecidable to determine whether any $t'$ is a bound on $t$, let alone find an optimal bound;
(ii) in the case of real arithmetic, reasoning is often prohibitively expensive.
In the setting of finding bounds, the challenge is finding a finite object to represent the infinite set of upper-bounds implied by the formula $\phi$.
In the case of \emph{linear} arithmetic, convex polyhedra can be represented finitely, and can completely represent the set of inequality consequences of a linear formula $\phi$.
Moreover, manipulating polyhedra is often reasonably efficient. However, in the non-linear rational case no such complete object exists, and in the non-linear real case manipulating the corresponding object\footnote{See the discussion on Positivestellensatz in \sectref{postivestull}.} is computationally challenging.

To address this challenge we introduce a mathematical object we call a \emph{cone of polynomials} (\sectref{coneRed})
to hold on to an infinite set of non-linear inequalities.
A cone of polynomials, consists of a polynomial ideal (\sectref{ideals}), which captures equations, and a polyhedral cone (\sectref{linearCones}), which captures inequalities.
Cones of polynomials strike a balance between expressiveness and computational feasibility---using \emph{non}-linear reasoning on \emph{equalities} through the ideal, and \emph{linear} reasoning on \emph{in}equalities through the linear cone, gives efficient yet powerful \emph{non}-linear reasoning on \emph{in}equalities.

We utilize cones of polynomials to address the non-linear OSB problem in a two-step process: (1) From $\phi$ create an implied cone of polynomials $C$. That is, $C$ is an infinite collection of inequalities, each implied by $\phi$. (2) \emph{Reduce} the term $t$ by $C$ to obtain $t^{*}$.

Due to the difficulties of non-linear arithmetic the first step is necessarily incomplete. However, the second step (reduction) is \emph{complete} with respect to cones of polynomials. Our reduction method for cones of polynomials makes use of a sub-algorithm that reduces a linear term by a polyhedron. That is, in \sectref{polyhedronRed} we give an algorithm (\algref{polyReduce}) that \emph{solves} the OSB problem, where $\phi$ is a conjunction of linear inequalities (a polyhedron), $\models$ is interpreted as linear arithmetic, and $\preceq$ is an order that encodes preferability of the dimensions of the returned bound.
This method makes use of a novel \emph{local projection} (\sectref{polyhedronRed}) method. Local projection can be seen as an incomplete method of quantifier elimination for polyhedra that avoids the blow-up of a full-projection method such as Fourier-Motzkin. Nevertheless, local projection suffices in the case of the OSB problem for polyhedra. In \sectref{experiments}, we compare our reduction method based on local project with a, perhaps more obvious, approach based on multi-objective linear programming. We find that our algorithm solves the problem much more efficiently than the LP approach.

In \sectref{coneRed}, we show (\theoref{conePRed}) how the polyheral OSB solution can be extended to the setting of cones of polynomials. This means that in particular we are able to completely solve OSB with respect to a polynomial $t$ and a polynomial cone $C$, which has the property that the result, $t^{*}$, is optimal with respect to any other bound $s$ implied by the cone $C$. This method works for desirability orders $\preceq$ that are \emph{monomials orders}.

With these methods in hand, \sectref{saturation} shifts to the following problem: Given a formula $\phi$, extract an implied cone $C$ for which the methods from \sectref{reduction} can be applied. 
Due to the issues of non-linear arithmetic, such an extraction process will necessarily be incomplete. However, in \sectref{saturation} we give a heuristic method for extracting a cone from a non-linear formula that works well in practice (\sectref{experiments}).
Moreover, our method allows $\phi$ and $t$ to contain function symbols, such as $\floor{\cdot}$ and $\frac{1}{\cdot}$ (reciprocal), both of which are outside the signature of polynomial arithmetic.
Overall, our two-step method is sound with respect to non-linear arithmetic
augmented
with additional function symbols. 

In \sectref{effectiveDeg}, we introduce the \emph{effective-degree order}.
Effective-degree is essentially a degree monomial orders, extended to the case of non-polynomial function symbols.
Effective-degree orders capture the intuitive notion that terms with fewer products are simpler.
Also, variable restriction can be encoded as an effective-degree order.

Our saturation and reduction methods combined with effective degree results in a powerful yet practical method for addressing the non-linear OSB problem.
In \sectref{experiments}, we give experimental results that show our method, 
using effective-degree as a term desirability order, produces interesting and relevant
bounds using a set of benchmarks extracted from Solidity code by industry experts in smart-contract verification.
Our tool is able to produce in seconds or minutes bounds which match or nearly-match human-produced bounds, as well as bounds where ones were previously unknown to human experts.

\paragraph{Contributions}
\begin{enumerate}
    \item The introduction of the optimal symbolic-bound synthesis problem
    \item The local-projection method for projecting polyhedra (\sectref{localProj})
    \item Algorithms for reducing a term $t$ by a polyhedron (\sectref{polyhedronRed}) and a polynomial cone (\sectref{coneRed})
    \item A saturation method that extracts a polynomial cone from a non-linear formula with additional function symbols (\sectref{saturation})
    \item The introduction of the effective-degree order on terms, which is amenable to automation, and in practice results in useful bounds (\sectref{effectiveDeg})
    \item An experimental evaluation demonstrating the power and practicality of our method (\sectref{experiments})
    %\john{Our tool matches or nearly matches human produced bounds on... Our tool produced bounds in X cases where a bound was unknown to a human} (\sectref{experiments})
\end{enumerate}
\sectref{related} discusses related work. Proofs of theorems are given in 
\iflong
appendices.
\else
supplementary material.
\fi

%% file: overview.tex
\section{Overview}\label{Se:overview}
\begin{figure}[t]
  \begin{subfigure}{.32\textwidth}
     \begin{lstlisting}[style=base, basicstyle=\small]
class Rebase {
  uint base
  uint elastic
  function add(amount) {
    a2base = toBase(amount)
    elastic += amount
    base += a2base
  }
  function toBase(x) {
    x * this.base / this.elastic
  }
}
\end{lstlisting}
     \caption{Extracted Solidity code.}
     \label{Fi:elastic}
    \end{subfigure}
\begin{subfigure}{.66\textwidth}
\centering
   \input{figures/overviewfig}
   \caption{Overview of the method}
   \label{Fi:overview}
\end{subfigure}
\end{figure}
% https://github.com/boringcrypto/BoringSolidity/blob/master/contracts/libraries/BoringRebase.sol
To motivate the optimal symbolic-bound synthesis problem, as well as understand how we address it, consider the code in \figref{elastic}. 
This code presents us with an interesting non-linear inequational-reasoning problem, which arises from a common smart-contract pattern.
A typical ``rebase'' or ``elastic'' smart contract holds some amount of ``tokens,'' which can vary over time, and each user holds a certain amount of ``shares'' in the tokens. While the number of tokens may vary (e.g., to control the price), the given number of shares that the user holds should correspond to a largely-unchanging percentage of the total tokens.
% https://www.coindesk.com/learn/what-is-a-rebaseelastic-token/
% It is based on a Solidity smart contract code for a simple ``rebase'' or ``elastic'' token, whereby the number of available ``tokens'' can be updated (to control the price), but the percentage of tokens that each user possesses should remain the same. The percentage a user owns is represented as a number of ``shares'' out of the total number of tokens.
The utility class \cinline{Rebase}, which is based on real-world Solidity code\footnote{\url{https://github.com/sushiswap/BoringSolidity/blob/master/contracts/libraries/BoringRebase.sol}}, tracks the total number of tokens in \cinline{elastic} and the total number of shares in \cinline{base}. The function \cinline{add} increases the number of tokens, and the number of available shares accordingly.
However, a given amount $v$ should be represented by the same number of shares even after an \cinline{add} operation.
Thus,
for a given values $v$ and $a$, 
%we are interested in how executing \cinline{add(a)} can change the value of \cinline{toBase(v)}.  That is, 
if we execute the sequence
\begin{center}
    \cinline{x = toBase(v); add(a); y = toBase(v)},
\end{center}
the term $t = x-y$ should be $0$, or close to $0$. 
Plugging in concrete value shows that $t$ is not identically $0$, but how far from $0$ is it?
The answer can depend on $v$ and $a$, as well as the initial values of \cinline{elastic}, and \cinline{base}, so a precise characterization involves a \emph{symbolic} expression.
Indeed, a verification expert that analyzed this problem came up with the bound $t \leq 1 + \floor{v/(e+a)}$, where $e$ is the initial value of \cinline{elastic}.
Can we automate this creative process of \emph{generating} a bound, which
for humans
often involves much trial-and-error, while even \emph{validating} a guess for a bound is challenging?
In this case, can we automatically find lower and upper symbolic bounds on the term \cinline{t = x-y}?

The same question can be translated 
to an OSB problem
by writing
the following conjunctive formula that represents the assumptions about the initial state, together with the program's execution:
\[\phi \defeq x = \floor{\frac{vb}{e}}\land y = \floor{\frac{vb'}{e'}} \land a2 = \floor{\frac{ab}{e}} \land e' = e +a \land b' = b + a2 \land a, b, e, v \ge 0.\]
The goal is to produce a term $t^{*}$ such that $\phi\models x-y\le t^{*}$.
Furthermore, we are interested in terms that are ``insightful'' in some sense. For example, we would like to produce a bound that does not contain any temporary variables $e'$, $a2$, or $b'$, as well as the variables $x$ and $y$. This variable restriction does not alone determine a desirable bound, but for this example we at least require a bound to satisfy this constraint. For this example, as well as our experiments, we use the effective-degree order (\sectref{effectiveDeg}) as a stand-in for term desirability. Using effective-degree we can encode variable restriction into the order, ensuring that the variables $e'$, $a2$, $b'$, $x$, and $y$ are absent from the bound we produce if such a bound exists. Effective-degree goes further and roughly minimizes the number of products in the result.

\figref{overview} gives an outline of our method.
% As stated in \sectref{introduction}, our method for solving this instance of OSB consists of a two-step process. First, we must construct an implied polynomial cone from this formula. Then, we must reduce the term of interest by the resulting cone. \figref{overview} outlines the process of producing a cone. More details about the saturation step can be found in \sectref{saturation}.
% \twr{Why this comment about the saturation step here?
% We should explain the elements of \figref{overview} first.
% }
% We note that in this overview section as well as \figref{overview}, we consider an input formula $\phi$ of the form $E\land I\land \phi$, where $E$ is a conjunction of equalities and $I$ is a conjunction of inequalities. That is, the equality and inequality assumptions have been separated out by the user; however, $\phi$ need not be of this form for our method to work. If the $E$ and $I$ assumptions were given within some clause of $\phi$ and indeed were implied by $\phi$ we would extract them later during the saturation step.
The first step to produce an implied cone is to \emph{purify} the formula $\phi$ into a formula using only polynomial symbols. We do this by introducing a new variable for each non-polynomial function, and placing the variable assignment in a \emph{foreign-function map}. 
That is, for every non-polynomial function symbol $f(w)$ we introduce a new variable, $u$, add $u\mapsto f(w)$ to our map and replace $f(w)$ with $u$. Purifying the formula $\phi$ we obtain,
\begin{align*}
    \phi' \defeq\ &x = u_4 \land y = u_5\land a2 = u_6 \land e' = e+a \land b' = b + a2 \land a,b,e,v \ge 0\\
    TM = &\{u_1\mapsto e^{-1}, u_2 \mapsto e'^{-1}, u_3 \mapsto e^{-1}, u_4 \mapsto \floor{vbu_1}, u_5 \mapsto \floor{vb'u_2}, u_6\mapsto \floor{abu_3}\}
\end{align*}

The purpose of purification is to produce a $\phi'$ which contains no function symbols. The original formula $\phi$ is equivalent to $\phi' \land \bigwedge_{u\mapsto f(w)} u = f(w)$. By making this separation of $\phi$ in $\phi'$ and $TM$ we can create methods that can separately work on $\phi'$, $TM$, or the combination of $\phi'$ and $TM$ when required. We then %can
strengthen $\phi'$ by adding properties of the functions in $TM$. This is the Instantiate Axioms step in \figref{overview}.\footnote{Formally we may consider the axioms for function symbols in the formula $\phi$ to be given in $\phi$. However, for convenience, our system automatically instantiates the mentioned $\floor{-}$ and $\frac{1}{-}$ axioms}
For example, $u_4$ represents a floor term and so satisfies more properties than a generic polynomial variable. Thus, after purification our method uses
the term map $TM$ to instantiate axioms
for floor and inverse for each occurrence of a floor and inverse term appearing in the map, and adds them to $\phi'$.
That is, we create $\phi''$ by adding the instantiated axioms $(u_1e = 1)$\footnote{On its own $u_1e=1$ is not a sound axiom, due to division by zero issues. However, in our applications (\sectref{experiments}) it can be assumed this division by zero never happens, thanks to SafeMath libraries in Solidity. 
%this is not a problem in our application We assume, due to safe math operations in Solidity, division by zero is not possible. 
A more generally-applicable axiom would be $e=0\lor u_1e=1$ (and explicitly assuming $e\neq 0$ in the  input), which our system can likewise handle.
%, which implies $u_1e=1$. 
%Our system is capable handling the more general axioms, but for brevity we use the axiom here with the assumption division by zero is not possible.
}, $(u_2 e' = 1)$, \dots, $vbu_1-1\le u_4\le vbu_1$, $vb'u_3-1\le u_5\le vb'u_3$, \dots, $e\ge 0\implies u_1\ge 0$, $vbu_1\ge 0\implies u_4\ge 0$, etc, to $\phi'$. At this point $\phi''$ is
\begin{align*}
    \phi''\defeq\ \phi' \land &(e\ge 0\implies u_1\ge 0) \land \dots \land (eu_1=1)\land \dots \land (vbu_1-u_4\ge 0) \land \dots
\end{align*}

After axioms have been instantiated, $\phi''$ and the term map $TM$ are used to construct a \emph{cone of polynomials} (\sectref{coneRed}). A cone of polynomials is a composite of a \emph{polynomial ideal} (\sectref{ideals}) and a \emph{polyhedral cone} (\sectref{linearCones}). The ideal and polyhedral cone are each represented by a finite set of basis equations and inequalities, respectively.
The ideal consists of its basis equations, as well as all other equations that are \emph{polynomially implied} by the basis equations. That is, the ideal consists of polynomials $p_1,\dots,p_k$, representing assumptions $p_i=0$, as well as any polynomial of the form $h_1p_1+\dots+h_kp_k$ for polynomials $h_1,\dots,h_k$.
The polyhedral cone
consists of its basis inequalities as well as all other inequalities that are \emph{linearly implied} by the basis inequalities. That is, the polyhedron consists of polynomials $q_1,\dots, q_r$, representing assumptions $q_i\ge 0$, as well as any other polynomial of the form $\lambda_1q_1+\dots \lambda_rq_r$ for scalar $\lambda_i\ge 0$. Overall, the cone consists of terms of the form $p+q$ where $p$ is a member of the ideal and $q$ is a member of the polyhedron. Because $p$ is an implied equation and $q$ is an implied inequality, we have $p+q\ge 0$.

We call the process of creating a cone of polynomials from $\phi''$ and $TM$ \emph{saturation}. We describe the saturation process in \sectref{saturation} using the running example from this section. At a high level, saturation is an iterative process that extracts equalities and inequalities that are implied by $\phi''$ and $TM$. A cone of polynomials is created by adding extracted equalities to the ideal part and by adding extracted inequalities to the polyhedral cone part. The methods that we use include congruence closure (\sectref{closure}), linear consequence finding (\sectref{consFinding}), and ``taking products'' (\sectref{takingProducts}). By taking products, we mean that from an inequality $w\ge0$ and $z\ge 0$, we can derive $w^2\ge0$, $wz\ge0$, $z^2\ge0$, etc. There are an infinite set of products we could add so our method takes products up to a given \emph{saturation depth}. In our experiments a saturation depth of 3 worked well. By bounding the set of products we add as well as the use of our consequence finding method makes saturation incomplete for full non-linear arithmetic. However, our experiments show saturation works well in practice (\sectref{experiments}).

As detailed in~\sectref{saturation}, saturation produces the following cone of polynomials on the running example:
\begin{align*}
    C \defeq\ &\langle x-u_4,y-u_5, a2-u_6, e'-e-a, b'-b-a2, u_3-u_1, eu_1 - 1, (e+a)u_2 - 1\rangle + \\
              &[e, a, v, b, e^2, ea, \dots,vbu_1 - u_4, vu_2u_6, vbu_2 - vbu_1+vu_2,u_5-vu_2u_6 - vbu_2 + 1,  1];
\end{align*}
$\langle x-u_4,\dots, \rangle$ is the ideal and $[e,\dots, 1]$ is the polyhedral cone. In other words, saturation extracted the equations $x-u_4 = 0$, $y-u_5=0$, etc. as well as the inequalities $e\ge 0$, $a\ge 0$, and many more.\footnote{For this example, saturation extracted 814 inequalities.}

The next step in addressing the OSB problem is to \emph{reduce} our term of interest by $C$.
That is, we need to find the best $t^{*}$ such that the cone implies $t\le t^{*}$.
Equivalently, the problem is to find the best $t^{*}$ such that the cone contains $t^{*}-t$.
Our reduction procedure for polynomials and polynomial cones works by first reducing the polynomial of interest by the ideal, and then reducing the result by the polyhedron.
The process of reducing the polynomial by the ideal is a standard method in computational algebraic geometry (\sectref{ideals});
however, we present a novel polyhedral reduction method (\sectref{polyhedronRed}), which in turn uses a novel projection method (\sectref{localProj}).

The main idea of our polyhedral reduction method is to order the dimensions of the polyhedron, which in our setting correspond to monomials, and successively \emph{project} out the worst dimension of the polyhedron\footnote{
  For example, any dimension that corresponds to a monomial that involves the unwanted variables $e'$, $a2$, or $b'$.
}
until the term of interest $t$ becomes unbounded.
We show (\theoref{polyReduce}) that the bound on $t$ right before $t$ becomes unbounded is optimal in the order of dimensions. In \sectref{coneRed}, we show that the combination of the standard ideal reduction with the polyhedral reduction yields a reduction method for the combined cone.

% Our methods do not address the OSB problem with a generic desirability order.
% Working backwards, the polyhedral reduction method works for orders that order the dimensions of the polyhedron. 
% In our setting, dimensions of the polyhedron correspond to monomials.
% In turn, the method for reducing a polynomial requires a \emph{monomial order}, which is a standard definition from computational algebraic geometry (\sectref{ideals}).
% Our equality-closure process also makes use of the method of reducing a polynomial by an ideal, but has additional restrictions on how monomials that correspond to function terms compare with other monomials.
% We summarize these restrictions with the \emph{effective-degree order} (\sectref{effectiveDeg}).
% Effective degree can be used to enforce a variable-restriction requirement, as well as work to minimize the number of multiplications in polynomials.
For the example from \figref{elastic}, we instantiate an effective-degree order that favors terms without temporary variables. From the saturated cone of polynomials $C$, we have the following equations in the basis of the ideal, $x-u_4$ and $y-u_5$, as well as the following inequalities in the basis of the polyhedral cone:
\[
  vbu_1 - u_4 \ge 0 \qquad va2u_2 + vbu_2 - vbu_1 + vu_2\ge 0 \qquad u_5 - va2u_2 - vbu_2 + 1\ge 0
\]
Reducing $x-y$ by the equations yields $u_4 - u_5$.
The polyhedral reduction method can then be seen as rewriting $u_4-u_5$ to $vu_2 + 1$ via the justification
\begin{align*}
    vu_2+1 - (u_4 - u_5) = vbu_1 - u_4+va2u_2 + vbu_2 - vbu_1 + vu_2+u_5 - va2u_2 - vbu_2 + 1
\end{align*}
The right-hand-side is non-negative.
Thus, $x-y\le vu_2+1$.
Before returning the final result, our system \emph{unpurifies} this bound by replacing $u_2$ with its definition in $TM$.
Consequently, our system returns the final result as ``$x-y\le \frac{v}{a+e} + 1$.''
Our system can also be used to automatically find a lower bound for a term.
In our example, the lower bound that it finds is ``$x-y\ge -1$.''

\label{Se:UnderFloorRewrite}
These bounds, $-1\le x-y\le \frac{v}{a+e}+1$, which the implementation of our method found on the order of seconds, are very nearly the bounds, $0\le x-y\le \floor{\frac{v}{a+e}}+1$, found manually by a human analyst.
Differences between the bound we compute automatically and the bound produced by a human sometimes stem from slightly different preferences in the tension between the bound's simplicity and tightness, but in this case a deeper issue is at play.
Our method has a limited capacity to perform inequality reasoning \emph{inside} a floor term;
for instance, we do not produce the inequality $\floor{t_1} \leq \floor{t_2}$ even when $t_1 \leq t_2$ is known, 
if $t_1$ or $t_2$ are not present in the input formula.
%(we discuss this issue in detail in~\Cref{Rem:UnderFloorRewrite}.
%(This axiom is hard to instantiate systematically and automatically.
%\twr{``hard'' without what happening?  The system spiraling out of control generating more and more floor terms?}
%)
We do obtain the slightly weaker $\floor{t_1} \leq t_2$, which, for instance, does not precisely cancel with $-\floor{t_2}$, leading to slightly weaker bounds.

Our initial experience with the system is that it is able to produce interesting upper bounds that are challenging to come up with manually. In one case (fixed point integer arithmetic---see~\sectref{experiments}), we asked a human analyst to propose a bound for a problem they knew, but had previously attempted only a bound in the other direction (whereas our system computes both at the same time).
After approximately fifteen minutes, and correcting the derivation at least once, they came up with a bound that nearly matches the bound that our system generated in less than a second.

%% file: figures/overviewfig.tex
\begin{tikzpicture}[
  scale = 0.6,
  mynodes/.style = {very thick, draw=black!50,top color=white,bottom color=black!20, font = \ttfamily, minimum width = 3.4cm},
  ideal/.style = {mynodes, diamond, aspect=2, rounded corners},
  tmap/.style = {mynodes,rectangle, rounded corners=3mm},
  pset/.style = {mynodes, rectangle},
  eqR/.style = {rounded corners, draw, top color=blue!15, bottom color=blue!25},
  cone/.style = {rounded corners, draw, dashed, top color = red!15, bottom color = red!25},
  textnode/.style = {inner sep = 1 pt, align=center,font=\small},
]

\def\hdist{5.5cm}

\node (phi) [textnode] {$\phi$};
%\node (ineqA) [below=0cm of eqA, textnode] {Ineq assums};
%\node (implA) [below=0cm of ineqA, textnode] {Other assums};
\node (ts) [above = 0cm of phi, textnode] {Terms to rewrite};

%\node(inputs) [fit=(eqA) (ineqA) (implA) (ts), draw, dashed, rounded corners, label={above:Input}] {};
\node(inputs) [fit=(phi) (ts), draw, dashed, rounded corners, label={above:\small Input}] {};

\node(tsP) [textnode, right=\hdist of ts.center, anchor = center] {Terms to rewrite};
\node(phiP) [textnode, right=\hdist of phi.center, anchor = center] {$\phi'$};
%\node(ineqP) [textnode, right=\hdist of ineqA.center, anchor = center] {Ineq assums};
%\node(implP) [textnode, right=\hdist of implA.center, anchor = center] {Other assums};
%\node(inputsP) [fit = (eqP) (ineqP) (implP) (tsP), draw, dashed, rounded corners, minimum width = 1.8cm, label={above:Polynomials}] {};
\node(inputsP) [fit = (phiP) (tsP), draw, dashed, rounded corners, minimum width = 1.8cm, label={above:\small Polynomials}] {};

\path (inputs) -- (inputsP) node [midway,textnode] (purify) {Purify};
\node(tmap) [textnode, below=0.66cm of purify] {Term map};

\node(instantiate) [textnode, below = 0.5cm of tmap] {Instantiate Axioms};

\draw[->] (tmap.south) -- (instantiate.north);

\draw [->] (inputs.east) -- (purify.west);
\draw [->] (purify.east) -- (inputsP);
\draw [->] (purify.south) -- (tmap.north);

\draw (instantiate.east -| phiP.south) node(axioms) [textnode] {$Ax \land \phi'$};
%\node(eqAx) [textnode, below =0cm of ineqAx] {Eq axioms};
%\node(implAx) [textnode, above =0cm of ineqAx] {Other axioms};
%\node (axioms) [fit = (eqAx) (ineqAx) (implAx), draw, dashed, rounded corners, label={above:Axioms}] {};
\node (assums) [fit = (axioms), draw, dashed, rounded corners, label={above:\small Assums + Axioms}] {};

\draw[->] (phiP.east) -- ++(right:2) node(mid) {} -- (\currentcoordinate |- axioms.east) -- (axioms.east);
%\draw[->] (phiP.east) -- (phiP.east -| mid.south) -- ++(right:5pt) -- (\currentcoordinate |- axioms.east) -- (axioms.east);
%\path (implP.east) -- (implP.east -| mid.south) -- ++(left:5pt) coordinate(lbit);
%\draw[->] (implP.east) -- (lbit) -- (\currentcoordinate |- implAx.east) -- (implAx.east);

%\draw[->] (instantiate.east) -- (implAx.west);
\draw[->] (instantiate.east) -- (axioms.west) node[midway] (midIA) {};
%\draw[->] (instantiate.east) -- (eqAx.west);

\node(satC) [textnode, below = 0.75cm of midIA] {Make and \\Saturate Cone};

\draw (tmap.west) -- ++(left:1.3cm) -- ++(down:2.3cm) coordinate(lctmap) -- (\currentcoordinate -| assums.south) -- (assums.south);
\draw [->] (lctmap -| satC.north) -- (satC.north);

\node(cone) [textnode, below = 0.75cm of satC] {Cone};

\draw[->] (satC.south) -- (cone.north);

\draw[->] (tsP.east) -- (tsP.east -| mid.south) -- ++(right:10pt) -- (\currentcoordinate |- cone.east) -- node[midway,label={above:\small Reduce}] {} (cone.east);

\node(res) [textnode] at (cone.east -| inputs.south) {Rewritten terms};

\draw[->] (cone.west) -- (res.east);

\end{tikzpicture}

%% file: background.tex
\section{Background}\label{Se:background}

Our method is based on the construction and manipulation of a cone, which as stated consists of a polynomial ideal to hold on to equations and a linear cone to hold on to inequalities. Part of our contribution is the use and manipulation of this composite object. However, we borrow many techniques and ideas from the study of the individual components. In this section, we give background on the definitions and properties of polynomials ideals and linear cones.

Overall, our method works for any ordered field. That is, our techniques are sound with respect to the theory of ordered fields. We will write $\models_{\textit{OF}}$ to denote entailment modulo the theory of ordered fields when we want to indicate soundness. Since $\mathbb{R}$ and $\mathbb{Q}$ are ordered fields, $\models_{OF}$ implies entailment with respect to non-linear real and non-linear rational arithmetic.

\subsection{Polynomials and Ideals}\label{Se:ideals}
In this section, we give definitions of polynomial ideals, as well as highlight algorithms and results that we will need in order to manipulate and reason about polynomial ideals.
For a more in-depth presentation of polynomial ideals and algorithms for manipulating them, see \citet{CLO:2015}.

Our use of the phrase monomial refers to the standard definition. We consider polynomials as being a finite linear combination of monomials over some (ordered) field $\mathbb{K}$. For example $\mathbb{K}$ could be $\mathbb{Q}$ or $\mathbb{R}$.

In this paper we use polynomial ideals to represent an infinite set of equality consequences. Due to some classical results from ring theory as well as a result due to Hilbert concerning polynomial ideals, we can take the following definition as a definition for any ideal of polynomials.
\begin{definition}
Let $\{p_1,\dots,p_k\}$ be a finite set of polynomials. $\langle p_1,\dots,p_k\rangle$ denotes the \emph{ideal generated by the basis $\{p_1,\dots,p_k\}$}. Furthermore, 
\[\langle p_1,\dots,p_k\rangle = \{h_1p_1+\dots + h_kp_k \mid 1\le i\le k \text{ and } h_i \text{ is an arbitrary polynomial}\}.\]
\end{definition}
If we consider a set of polynomials $p_1, \dots, p_n$ as a given set of polynomial equations, i.e., $p_i = 0$ for each $i$, then the ideal generated by $p_1, \dots, p_n$ consists of equational consequences.
\begin{example}\label{Exa:ideal}
 One way to see that $x - 1 - t = 0 \land y - 1 - t^2 = 0$  $\models_{\textit{OF}} x^2 - 2x = y - 2$, is by observing
    \begin{align*}
      x^2 - 2x - y + 2&\in \langle x-1-t, y - 1 - t^2\rangle\\
      x^2 - 2x - y + 2 &= (x - 1 + t)(x-1-t) + (-1)(y-1-t^2)
    \end{align*}
\end{example}

As will be highlighted shortly, determining ideal membership for $\mathbb{Q}[x_1,\dots, x_n]$, $\mathbb{R}[x_1,\dots, x_n]$ is \emph{decidable}.
However, in the general context of non-linear rational arithmetic or nonlinear integer arithmetic, determining polynomial consequences is \emph{undecidable}.
Therefore, in general ideals give a sound but incomplete characterization of equational consequences. A simple example illustrating this fact is $x^2=0\models_{\text{OF}} x = 0$, but $x\not\in \langle x^2\rangle$.

While polynomial ideals are in general incomplete, they have the advantage of having useful algorithms for manipulation and reasoning. Namely, the multivariate-polynomial division algorithm and the use of Gr\"obner bases give a practical method to \emph{reduce} a polynomial by an ideal and check ideal membership. These techniques are integral to our overall method, so we now briefly highlight some of these ideas.

Algorithms for manipulating polynomials often consider the monomials one at a time. Thus, we often orient polynomials using a \emph{monomial order}.
\begin{definition}\label{De:monomialOrder}
Let $\MOLE$ be a relation on monomials. $\MOLE$ is a \emph{monomial order}, if
\begin{enumerate}
    \item $\MOLE$ is a total order.
    \item For any monomials $a$, $b$, and $m$, if $a\MOLE b$, then $a\cdot m\MOLE b\cdot m$
    \item For any monomial $a$, $a\MOGE 1$.
\end{enumerate}
\end{definition}
With a monomial order defined, we often write polynomials as a sum of their monomials in decreasing order. A strategy for ensuring termination of algorithms is to process monomials in decreasing order with respect to some monomial order, while guaranteeing intermediate steps do not introduce larger monomials. Because a monomial order is well-ordered, termination is ensured.

Common monomial orders are the lexicographic monomial order, degree lexicographic order, and degree reverse lexicographic order (grevlex).
The details of these monomial orders in unimportant for this work, but in practice the grevlex order tends to yield the best performance in implementations.

\begin{definition}
With respect to a monomial order the \emph{leading monomial} of a polynomial $p$, denoted $\LM(p)$, is the greatest monomial of $p$.
\end{definition}

Once a monomial order has been defined, we can use the multivariate polynomial division algorithm to divide a polynomial $f$ by an ideal $\langle p_1,\dots, p_k\rangle$. This algorithm successively divides the terms of $f$ by the leading terms of the set of $p_i$'s until no more divisions can be performed. The result is a remainder $r$. The value of this remainder can be used for various purposes. For example, if $r=0$, then $f\in \langle p_1,\dots,p_k\rangle$. However, examples can be constructed that show that performing multivariate division on an \emph{arbitrary} basis does not necessarily yield a unique result. In other words, it is possible to have another basis $p'_1,\dots, p'_k$ with $\langle p'_1,\dots,p'_k\rangle = \langle p_1,\dots,p_k\rangle$, but dividing $f$ by $p'_1,\dots,p'_k$ will yield a different remainder $r'$. The solution to this issue is to divide by a \emph{Gr\"obner basis}. That is, to divide a polynomial $f$ by an ideal $\langle p_1,\dots,p_k\rangle$, we do not divide $f$ by $p_1,\dots,p_k$, but instead we construct a Gr\"obner basis $g_1,\dots,g_s$ with $\langle g_1,\dots,g_s\rangle = \langle p_1,\dots,p_k\rangle$. It can then be shown that dividing $f$ by $g_1,\dots,g_s$ will yield a unique remainder.

The exact definition of a Gr\"obner basis is technical and not required for this paper.
What is required to know is that, given an ideal $\langle p_1,\dots,p_k\rangle$, there are algorithms, such as Buchberger's algorithm \cite{buchbergerA,buchbergerB} and the F4 \cite{FaugereF4} and F5 \cite{FaugereF5} algorithms, for constructing a Gr\"obner basis $g_1,\dots,g_s$ with $\langle g_1,\dots,g_s\rangle = \langle p_1,\dots, p_k\rangle$.
Furthermore, using the multivariate division algorithm to divide a polynomial $f$ by a Gr\"obner basis yields a remainder with certain special properties.
\begin{definition}
Let $B$ be a Gr\"obner basis, and $f$ a polynomial. We call the process of dividing $f$ by $B$ using the multivariate division algorithm and taking the remainder \emph{reduction}, and denote the process by $\textbf{red}_B(f)$.
\end{definition}
\begin{theorem}\label{The:grobnerBasis}\cite[][Section 2.6]{CLO:2015}
Let $B = \{g_1,\dots,g_s\}$ be a Gr\"obner basis for an ideal $I$, $f$ a polynomial, and $r = \textbf{red}_B(f)$. $r$ is the \emph{unique} polynomial with the following properties:
\begin{enumerate}
    \item No term of $r$ is divisible by any of $\LT(g_1),\dots,\LT(g_s)$.
    \item There is a $g\in I$ with $f = g+r$.
\end{enumerate}
\end{theorem}
\begin{lemma}\label{Lem:reduceReduces}
$\LT(f)\MOGE \LT(\textbf{red}_B(f))$ for any $f$ and basis $B$.
\end{lemma}

\begin{corollaryrep}
    Let $B = \{g_1,\dots,g_s\}$ be a Gr\"obner basis for an ideal $I$, $f$ a polynomial, and $r = \textbf{red}_B(f)$. $r$ is the optimal remainder in the monomial order. That is, for any other $r'$ with $f=g'+r'$ for some $g'\in I$, $\LT(r')\MOGE \LT(r)$.
\end{corollaryrep}
\begin{appendixproof}
Let $r'$ be some polynomial with $f=g'+r'$. Reducing $r'$ by $B$ yields $r''$ with $r'=g''+r''$ for some $g''\in I$, as well as $\LT(r')\MOGE \LT(r'')$. Moreover, $r''$ satisfies the properties in \theoref{grobnerBasis} with respect to $r'$. However, $f = g'+g''+r''$, and because $g', g''\in I$ so is $g'+g''\in I$. Thus, $r''$ also satisfies the conditions of \theoref{grobnerBasis} with respect to $f$. Thus, $r = r''$ and $\LT(r')\MOGE \LT(r'') = \LT(r)$.
\end{appendixproof}
\begin{corollary}\label{Cor:idealMem}
    Let $B = \{g_1,\dots,g_s\}$ be a Gr\"obner basis for an ideal $I$, $f$ a polynomial. $f\in I$ if and only if $0 = \textbf{red}_B(f)$.
\end{corollary}

\subsection{Polyhedral Cones}\label{Se:linearCones}
In this section, we give background on polyhedral cones.
Mirroring the process of using ideals to represent equations, we use polyhedral cones to represent inequalities. 
The reader should keep in mind that our method uses two different kinds of cones.
We have an inner cone which is used to hold on to linear inequalities and an outer cone which consists of an ideal and the inner cone. The inner cone is a \emph{polyhedral cone} and is the main subject on this section. 
We will describe the outer cone in more detail in \sectref{coneRed}.
To make the distinction between the two concepts clear, we will use the terms ``polyhedral cone'' and ``cone of polynomials'' to refer to the inner and outer cone, respectively.

\begin{definition}\label{De:linearCone}
Let $\mathbb{K}$ be an ordered field (e.g., $\mathbb{R}$ or $\mathbb{Q}$) and $V$ be a vector space over $\mathbb{K}$. A \emph{polyhedral cone} $C$ is the conic combination of finitely many vectors. That is, there is a set of vectors $\{v_1,\dots,v_n\}$ with $C = \{\lambda_1v_1+\dots +\lambda_nv_n\mid \lambda_i\ge 0\}$. We use $C=[v_1,\dots,v_n]$ to denote that $C$ is generated by the vectors $\{v_1,\dots,v_n\}$.
\end{definition}

While we use polyhedral cones to \emph{represent} a set of linear consequences, we frame some of our reduction algorithms (\sectref{polyhedronRed}) in terms of \emph{convex polyhedra}. 
Fortunately, there is a very strong connection between polyhedral cones and convex polyhedra.
There are multiple equivalent definitions for a convex polyhedron that lead to different representations.
In this paper we only %consider the case where we 
represent a polyhedron using a set of inequality constraints, sometimes called the \emph{constraint representation}.

\begin{definition}
Let $\mathbb{K}$ be an ordered field.
A \emph{linear constraint} over variables $x_1,\dots,x_n$ is of the form $a_1x_1+\dots+a_nx_n + b\ge 0$, $a_1x_1+\dots+a_nx_n + b> 0$, or $a_1x_1+\dots+a_nx_n + b = 0$, where $a_1,\dots,a_n,b\in \mathbb{K}$. A (convex) \emph{polyhedron} is the set of points of $\mathbb{K}^n$ satisfying a set of linear constraints.
\end{definition}

Because each equality can be represented as two inequalities, we could consider polyhedra to not have equality constraints.
However, having explicit equalities can allow algorithms to be more efficient in their calculations.
We do not take a strong stance on whether all of the equalities of a polyhedron are explicit or not.
In \sectref{reduction}, we sometimes consider equality constraints as being explicitly part of a polyhedron, but our methods work the same if the equalities are implicit.

If we look at the constraints of the polyhedron as given inequality assumptions, taking conical combinations of the constraints give a sound set of inequality consequences. Moreover, Farkas' Lemma shows that this set of consequences is also \emph{complete}.
\begin{lemma}\label{Lem:linComp}(Variant of Farkas' Lemma)
Let $P$ be a non-empty polyhedron with non-strict constraints $N=\{c_1\ge 0, \dots, c_k\ge0\}$ and strict constraints $S=\{s_1>0,\dots, s_l>0\}$. Let $P^{*}$ denote the polyhedral cone $[c_1, \dots, c_k, s_1,\dots,s_l, \mathbf{1}]$. Then $P\models_{\textit{LRA}} t\ge 0$ if and only if $t \in P^{*}$.
\end{lemma}

We close this section by giving observing that a polyhedral cone that represents inequalities can also represent
equalities
in the sense that for some vector $v\in C$ it is possible for $-v\in C$ to hold as well.
If $C$ is holding onto non-negative vectors, we would have $v\ge 0$ and $-v\ge0$, so $v=0$.
\begin{definition}\label{De:salient}
If the only vector $v$ of $C$ with $-v\in C$ is $\mathbf{0}$ then $C$ is called \emph{salient}.
\end{definition}
In the case of polyhedral cones we can determine if a cone is salient by looking at the generators.
\begin{lemmarep}\label{Lem:salient}
If $C=[v_1,\dots,v_n]$ is not salient then there is an $i\in \{1,\dots,n\}$ with $v_i\in \{v_1,\dots,v_n\}$ and $-v_i\in C$.
\end{lemmarep}
\begin{appendixproof}
Let $C$ be a non-salient cone. Then there exists a vector $f\in C$ and $-f\in C$. Thus $f=\lambda_1v_1+\dots+\lambda_n v_n$ and $-f = \lambda'_1v_1+\dots + \lambda'_nv_n$. There must be some $\lambda'_i$ non-zero. Without loss of generality let $\lambda'_1>0$. Then $\lambda'_1(-v_1) = f + \lambda'_2v_2+\dots+\lambda'_nv_n$. Substituting the equation for $f$ we have $\lambda'_1(-v_1) = \lambda_1v_1+\dots+\lambda_nv_n + \lambda'_2v_2+\dots+\lambda'_nv_n$. Dividing by $\lambda'_1$ gives $-v_1 = \frac{\lambda_1}{\lambda'_1}v_1 + \frac{\lambda_2+\lambda'_2}{\lambda'_1}v_2 + \dots + \frac{\lambda_n+\lambda'_n}{\lambda'_1}v_n$. Thus $-v_1\in C$. Also $v_1\in C$.
\end{appendixproof}

\begin{remark}
\lemref{linComp} can be modified to also say something about the strict inequality consequences of $P$. However, we would have to add the condition that the ``witness'' of $t\in P^{*}$ has at least one of the multiples on the strict inequality constraints non-zero. Consequently the same machinery presented here can be used to hold on to and decide strict inequalities. We just need to make sure that we keep appropriate track of which constraints are strict and which ones are not. \Tool does distinguish between strict and non-strict inequalities, and so it is possible for the tool to return an bound $t^{*}$ and know that $t<t^{*}$ rather than $t\le t^{*}$. However, for presentation purposes in the subsequent sections we focus on the non-strict case.
\end{remark}

%% file: reduction.tex
\section{Reduction}\label{Se:reduction}
In this section we present our algorithm for efficiently reducing a term w.r.t.\ a cone of polynomials. We first present its key technical component, the algorithm for local projection (\sectref{localProj}), then explain how to use it to perform reduction w.r.t.\ an (ordinary) polyhedron (\sectref{polyhedronRed}), and finally extend it to operate w.r.t.\ the extra ideal to handle the more general case of a cone of polynomials (\sectref{coneRed}).

\subsection{Local Projection}\label{Se:localProj}
An important polyhedral operation is \emph{projection}.
Our reduction method uses a weaker projection operation, which we call \emph{local projection}.
We could use a standard polyhedral-projection operation such as Fourier-Motzkin elimination to yield the same result.
However, using full Fourier-Motzkin elimination to remove a single variable from a polyhedron of $n$ constraints can result in $\mathcal{O}(\frac{n^2}{4})$ constraints in the projection. Projecting out $d$ variables can result in
$\mathcal{O}(\left(\frac{n}{4}\right)^{2^d})$
constraints, although many are redundant.
The number of necessary constraints grows as a single exponential, at the expense of some additional work detecting redundant constraints at each step.
In contrast to this complexity, using local project to remove a single variable is in linear in time and space.
Thus, projecting out $d$ variables takes $\mathcal{O}(dn)$ time and space.
The caveat is that local projection only results in a subset of the
real projection result,
but, as we will show, the
real projection result
can be finitely covered by local projections.
In the worst case the number of partitions for projecting out a single variable is $\mathcal{O}(n^2)$, so local projection does not give a theoretical advantage compared to Fourier-Motzkin.
However, in our case, we often do not need to compute the 
full projection result.
Instead, we only require parts of it, and so using local projection gives us a \emph{lazy} method for computing objects of interest.
Local projection can also be understood as a method of model-based projection \cite[][Section 5]{KGC:2016}, specialized to the setting of polyhedra. \citet{KGC:2016} give a model-based projection for LRA based on the quantifier-elimination technique of \citet{LW:1993}. Thus, our specialization is very similar to these prior methods.

\begin{definition}
Let $P$ be a polyhedron with dimension $d_i$. $\proj{P}{d_i} = \{m \models_{LRA} \exists d_i. P\}$.
\end{definition}

Let $P$ be a polyhedron represented by a conjunction of equality and inequality constraints, and let $x$ be a dimension of $P$. The constraints of $P$ can be divided and rewritten as follows:
\begin{itemize}
    \item Let $E_x \defeq \{x = -\frac{f_i}{e_i} \mid e_ix+f_i = 0 \in P\}$, where each $e_i$ is a constant and each $f_i$ is $x$-free.
    \item Let $L_x \defeq \{x \ge -\frac{b_i}{a_i} \mid a_ix+b_i\ge 0\in P\}$ and $L^s_x \defeq \{x > -\frac{b'_i}{a'_i} \mid a'_ix+b'_i\ge 0\in P\}$,  where each $a_i, a'_i$ is a constant greater than $0$, and each $b_i, b'_i$ is $x$-free.
    \item Let $U_x \defeq \{x \le -\frac{d_i}{c_i} \mid c_ix + d_i \ge 0\in P\}$ and $U^s_x\defeq \{x < -\frac{d'_i}{c'_i} \mid c'_ix + d'_i > 0 \in P\}$, where each $c_i, c'_i$ is a constant less than $0$, and each $d_i, d'_i$ is $x$-free.
    \item Other constraints, $C$, not involving $x$.
\end{itemize}

\paragraph{Local Projection} Let $x$ be some dimension of a polyhedron $P$ that is represented by equality constraints $E_x$, lower-bound constraints $L_x$ and $L^s_x$, upper-bound constraints $U_x$ and $U^s_x$, and other constraints $C$. Let $m$ be a model of $P$. The \emph{local projection} of $x$ from $P$ w.r.t. $m$, denoted by $\lproj{m}{P}{x}$, is a polyhedron defined by a set of constraints as follows:
\begin{itemize}
    \item If $E_x$ is not empty, then let $x=e\in E_x$. Then $\lproj{m}{P}{x}$ is
\[E_x[x\mapsto e] \cup L_x[x\mapsto e] \cup L^s_x[x\mapsto e] \cup U_x[x\mapsto e] \cup U^s_x[x\mapsto e]\cup C.
\]
    \item If $E_x$, $L_x$, and $L^s_x$ are empty, then $\lproj{m}{P}{x} \defeq C$
    \item If $E_x$ is empty, but either $L_x$ or $L^s_x$ is non-empty, then let\\ $L'_x = \{lb \mid x \ge lb \in L_x\} \cup \{lb'\mid x > lb' \in L^s_x\}$. Let $lb^{*} \in \argmax_{L'_x} \textit{eval}(lb,m)$. 
    \begin{itemize}
        \item If $lb^{*}$ corresponds to a non-strict constraint, then $\lproj{m}{P}{x}$ is
    \begin{align*}
        \{lb\le lb^{*}\mid lb\in L'_x\} \cup \{lb^{*} \le ub\mid x\le ub \in U_x\}\cup \{lb^{*} < ub'\mid x < ub' \in U^s_x\} \cup C
    \end{align*}
        \item If $lb^{*}$ corresponds to a strict constraint, then $\lproj{m}{P}{x}$ is
    \begin{align*}
 \{lb\le lb^{*}\mid lb\in L'_x\} \cup \{lb^{*} < ub\mid x \le ub \in U_x\}\cup \{lb^{*} < ub'\mid x < ub' \in U^s_x\} \cup C
    \end{align*}
    \end{itemize}
\end{itemize}
The idea of a local projection is identical to a full projection except in local project we only consider the lower bound $lb^{*}$ that is binding with respect to the given model. In general there are other models with different lower bounds, so a full projection needs to consider these alternative cases. However, because there are only finitely many possible binding lower bounds, local project finitely covers the full project. These ideas are formally captured by \lemref{localProj}. For a more detailed comparison of local projection versus full projection see the proof of the lemma.

\begin{lemmarep}\label{Lem:localProj}
Let P, be a polyhedron. For a model $m\models P$ and a dimension $x$, the following are true:
\begin{enumerate}
    \item $m\models \lproj{m}{P}{x}$
    \item $\lproj{m}{P}{x} \models \proj{P}{x}$
    \item $\{\lproj{m}{P}{x} \mid m \models P\}$ is a finite set
\end{enumerate}
\end{lemmarep}
\begin{appendixproof}
We begin by first giving a logical account of $\proj{P}{x}$. That is, we consider different cases and construct a quantifier free formula $\psi$ and show it's equivalent to $\proj{P}{x}$. We then compare local projection with the result to show the lemma.

Let $E_x$, $L_x$, $L^s_x$, $U_x$, $U^s_x$, and $C$ be the constraints of $P$. $\proj{P}{x}$ is $\exists x. \phi$ where $\phi$ is a conjunction of all of the $E_x$, $L_x$, $L^s_x$, $U_x$, $U^s_x$, and $C$ constraints. To compute a quantifier-free formula equivalent to $\exists x. \phi$ we consider a few cases. In each case, to see that the resulting quantifier-free formula, say $\psi$, is equivalent to $\exists x. \phi$, it suffices to construct a value for $x$ from any model of $\psi$ such that $\phi$ is satisfied.

\paragraph{$E_x$ is not empty.} To create an equivalent quantifier-free formula, we  pick and rewrite some equality from $E_x$, say $x =e$. Then we substitute $e$ in for $x$ in all the constraints in $E_x$, $L_x$, and $U_x$. That is 
\[\psi \defeq \bigwedge E_x[x\mapsto e] \cup L_x[x\mapsto e] \cup L^s_x[x\mapsto e] \cup U_x[x\mapsto e] \cup U^s_x[x\mapsto e]\cup C.\]
To construct a satisfying model of $\phi$ from a model $m$ of $\psi$, we set $x = \textit{eval}(e, m)$, where $\textit{eval}(e, m)$ is $e$ evaluated at $m$.

\paragraph{$E_x$ is empty and $L_x\cup L^s_x$ is empty.} In this case, $x$ has no lower bound. That is, $x$ can get arbitrarily small. In this case, $\psi \defeq \bigwedge C$. To construct a satisfying model of $\phi$ from a model $m$ of $\psi$, we find the smallest upper-bound constraint, and ensure that is satisfied.
For example, we set $x = -1 + \min(\min_{x\le ub \in U_x} \textit{eval}(ub, m), \min_{x <ub' \in U^s_x} \textit{eval}(ub', m))$.

\paragraph{$E_x$ is empty and $L_x\cup L^s_x$ is not empty.} This is the interesting case, and where the quadratic blow-up comes from.
Let $L'_x = \{lb \mid x\ge lb \in L_x\} \cup \{lb'\mid lb' > 0 \in L^s_x\}$. 
Consider the set of models where $lb^{*}$ is the binding lower-bound. That is, $m$ is a model with $lb^{*} \in \argmax_{L'_x} \textit{eval}(lb, m)$. The set of models for which $lb^{*}$ is the binding lower-bound are models that satisfy $\textit{eval}(lb^{*}, m)\ge \textit{eval}(lb, m)$ for each $lb\in L'_x$. Furthermore, if $x\ge lb^{*}$ (or $x> lb^{*}$, if $lb^{*}$ corresponds an element of $L^s_x$), then all the constraints on $x$ are satisfied. So far, all we have done is restrict the scope to the set of models where $lb^{*}$ is the binding lower-bound. A formula that represents this case is
\[
\phi_{lb^{*}}\defeq (\bigwedge_{lb\in L'_x} lb\le lb^{*}) \land (\bigwedge_{x \le ub\in U_x} lb^{*} \le x \le ub) \land (\bigwedge_{x < ub'\in U^s_x} lb^{*}\le x < ub') \land (\bigwedge C).
\]
(The lower-bounds on $x$ in the above formula would be strict if $lb^{*}$ corresponds to a strict bound.) Restricting to the case of $lb^{*}$ being the binding lower bound, $\phi_{lb^{*}}$ is equivalent to $\phi$. Finally, we can simply drop $x$ from the above formula, but keep transitively implied constraints. That is, let
\[
\psi_{lb^{*}}\defeq (\bigwedge_{lb\in L'_x} lb\le lb^{*}) \land (\bigwedge_{x \le ub\in U_x} lb^{*} \le ub) \land (\bigwedge_{x < ub'\in U^s_x} lb^{*} < ub') \land (\bigwedge C).
\]
(Once again the middle inequality may be strict, depending on $-\frac{b^{*}}{a^{*}}$.) $\psi_{lb^{*}}$ is equivalent to $\exists x. \phi_{lb^{*}}$. To get a value for $x$ from a model $m'$ of $\psi_{lb^{*}}$, we set 
\[x = \frac{1}{2}(\textit{eval}(lb^{*}, m') + \min(\min_{x\le ub \in U_x} \textit{eval}(ub, m'), \min_{x < ub'\in U^s_x} \textit{eval}(ub', m')).\]
That is, $x$ is the mid-point of the most binding lower-bound and the most binding upper-bound. Thus, $x$ will satisfy all lower-bounds and all upper-bounds in $\phi_{lb^{*}}$.
All of this reasoning was under the assumption that a particular lower-bound $lb^{*}$ was binding. However, assuming no lower-bound is redundant means that any $lb\in L'_x$ could be a binding lower-bound. Therefore, to construct a quantifier-free formula $\proj{P}{x}$ we simply take a disjunction among all possible lower bounds.
\[
\exists x.\phi \equiv \bigvee_{lb\in L'_x} \psi_{lb}.
\]

From this presentation, it is clear how local projection compares with full projection.
In the cases when $P$ has an equality constraint involving $x$, or when $x$ has no lower-bounds in $P$, $\lproj{m}{P}{x}$ and $\proj{P}{x}$ are equivalent. Otherwise, $\lproj{m}{P}{x}$ exactly matches one disjunct of $\proj{P}{x}$ as presented above. Thus, $\lproj{m}{P}{x} \models \proj{P}{x}$. The number of disjuncts of $\proj{P}{x}$ is finite, so $\{\lproj{m}{P}{x} \mid m \models P\}$. Furthermore, every disjunct of $\proj{P}{x}$ is satisfied by some $m$.
\end{appendixproof}
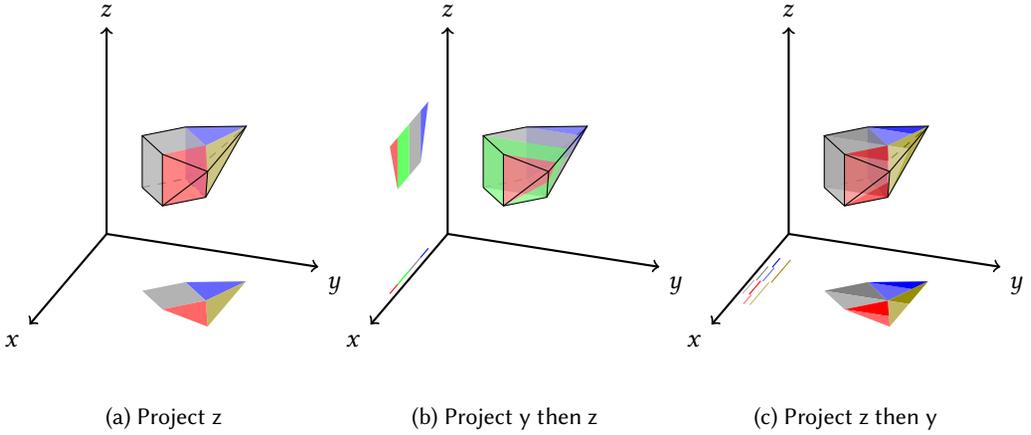
\begin{figure}[t]
   \begin{subfigure}{.32\textwidth}
     \input{figures/localprojfig1}
     \caption{Project z}
     \label{Fi:localProja}
    \end{subfigure}
    \begin{subfigure}{0.32\textwidth}
      \input{figures/localprojfig2}
      \caption{Project y then z}
      \label{Fi:localProjb}
    \end{subfigure}
    \begin{subfigure}{0.32\textwidth}
      \input{figures/localprojfig3}
      \caption{Project z then y}
      \label{Fi:localProjc}
    \end{subfigure}
    \caption{Local projections of a polyhedron.}
    \label{Fi:localProj}
\vspace{-0.2cm}
\end{figure}

\figref{localProj} gives a geometric picture of local projection and projection.
Consider \figref{localProja}, where the goal is to project out the $z$ dimension.
Take the red region for example.
Any model in the red region has the lower-front facing triangle as a binding constraint;
therefore, local projecting to the $x$-$y$ plane yields the red-triangle.
The union of the red, gray, olive, and blue regions give the full projection.
\figref{localProjb} is a similar diagram, but for projecting out $y$ then $z$.
\figref{localProjc} shows the result of projecting out $z$ then $y$.
The result is a line segment in the $x$ dimension.
In \figrefs{localProjb}{localProjc}, the resulting projections are depicted as being slightly displaced from the $x$-axis for clarity.

Local projection can also be used to project out multiple dimensions by projecting out each dimension sequentially. 
\begin{definition}
Given a list of dimensions, we use $\lproj{m}{P}{[d_1, \dots, d_k]}$ to denote\\ $\lproj{m}{\lproj{m}{\dots,\lproj{m}{P}{d_1},\dots}{d_{k-1}}}{d_k}$. Similarly, for $\proj{P}{[d_1,\dots, d_k]}$.
\end{definition}

Crucially, locally projecting out a set of variables has the same relationship to the full projection of the same variables
as we have in \lemref{localProj} for locally projecting out a single variable.

\begin{theoremrep}\label{The:localProjList}
Let P, be a polyhedron. For a $m\models P$ and a list of dimensions $[d_1,\dots,d_k]$, the following are true
\begin{enumerate}
    \item $m\models \lproj{m}{P}{[d_1,\dots,d_k]}$
    \item $\lproj{m}{P}{[d_1,\dots,d_k]} \models \proj{P}{[d_1,\dots,d_k]}$
    \item $\{\lproj{m}{P}{[d_1,\dots,d_k]} \mid m \models P\}$ is a finite set
\end{enumerate}
\end{theoremrep}
\begin{appendixproof}
These properties can be shown by induction on the length of the list of dimensions.
The base case is covered by \lemref{localProj}.
For the inductive step, suppose that the theorem holds for $\lproj{m}{P}{[d_1,\dots,d_{k-1}]}$.
By \lemref{localProj}, $\lproj{m}{P}{[d_1,\dots,d_{k-1}, d_{k}]}$, entails $\lproj{m}{P}{[d_1,\dots,d_{k-1}]}$, and finitely covers $\lproj{m}{P}{[d_1,\dots,d_{k-1}]}$.
Thus, by the inductive hypothesis, the theorem holds.
\end{appendixproof}

While it is well known that when performing a full projection the order in which the dimensions are presented does not matter;
however, in the case of local projection, the order \emph{does} matter.
To see why,
compare \figrefs{localProjb}{localProjc}.
In \figref{localProjb} the red, green, gray, and blue line segments are the possible results from projecting out y then z. However, in \figref{localProjc}, either of the olive line segments (the ones furthest from the axis) are possible results from projecting out z then y. There is no corresponding segment in \figref{localProjb} to either of the olive line segments.
However, \theoref{localProjList} ensures that the set of possible local projections is finite, and that they exactly cover the full projection.

\subsection{Polyhedral Reduction}\label{Se:polyhedronRed}
\begin{figure*}
   \begin{minipage}{0.46\textwidth}
   \begin{algorithm}[H]\small
\caption{Conjecture}\label{Alg:conjBound}
	\SetAlgoLined
	\KwIn{Polyhedron $P$, model $m$ of $P$, dimensions $d_1, \dots, d_k$ and $T$}
	\KwResult{Conjectured upper bounds on $T$}
	$P'\gets P$;
	$P''\gets \lproj{m}{P'}{d_1}$;
	$i\gets 2$\;
	\While{$T$ has an upper bound in $P''$}{
	  $P'\gets P''$\;
	  $P''\gets \lproj{m}{P''}{d_i}$\;
	  $i\gets i+1$\;
	}
	$B\gets \{\}$\;
	\For{$c\ge 0 \in P'$}{
	  \If{Coefficient of $T$ in $c$ is negative}{
	    $B\gets B \cup \{c\ge 0\}$\;
	  }
	}
    \Return $B$\;
\end{algorithm}
   \end{minipage}
\begin{minipage}{0.46\textwidth}
\begin{algorithm}[H]\small
\caption{Polyhedral reduction\label{Alg:polyReduce}}
	\SetAlgoLined
	\KwIn{Polyhedron $P$, term $t$ over variables $d'_1, \dots, d'_k$, order $\ll$}
	\KwResult{$r$ with $P\models t \le r$}
	$T \gets$ new dimension\;
	$P \gets P \cup \{T = t$\}\;
	$d_1, \dots, d_k\gets $ sort($\{d'_1, \dots, d'_k\}$, $\gg$)\;
	$U \gets \{\}$\;
	\While{$m\models P \land \bigwedge_{u\in U} \neg u$}{
	  $U\gets U\cup$ conjecture$(P, m, [d_1, \dots, d_k], T)$\;
	}
	$r \gets $ filter true bounds from $U$\;
    \Return $r$\;
\end{algorithm}
\end{minipage}
\vspace{-0.4cm}
\end{figure*}

In this section we present our algorithm for optimally reducing a polyhedron $P$ with respect to a linear term $t$ and an order on dimensions $\MOLE$. \algref{polyReduce} will produce a bound $t^{*}$ over dimensions $d_j,\dots,d_k$ such that $P\models_{LRA} t \le t^{*}$ and $t^{*}$ is optimal with respect to $\MOLE$. That is, for any $b$ with $\models_{LRA} t \le b$, then $b$ is an expression over the dimensions $d_i,\dots, d_k$ with $d_n\MOLE d_i$. Another way to think about the optimality is the ``leading dimension'' of $t^{*}$ is minimal.

\begin{wrapfigure}{R}{0.45\textwidth}
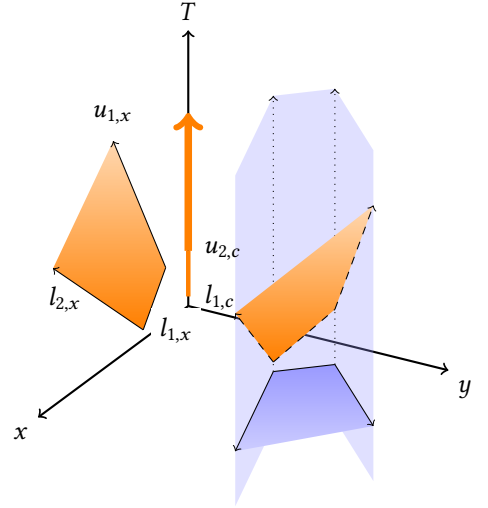

\vspace{-1cm}
\begin{minipage}{0.43\textwidth}
\begin{figure}[H]
    \centering
    \include{figures/localprojfig4}
    \vspace{-0.5cm}
    \caption{Reduction by local project}
    \label{Fi:localProjRed}
\end{figure}
\end{minipage}
\end{wrapfigure}
\figref{localProjRed} gives a geometric representation of \algref{polyReduce}.
Suppose that we wish to upper-bound some term $t$ that is an expression over $x$ and $y$, under the assumption that $x$ and $y$ are restricted to the (unbounded) dark-blue region. Let $T=t$. Then the floating orange plane is the term to optimize.
Suppose that we favor an upper-bound containing $x$ over an upper-bound containing $y$.
In \figref{localProjRed}, the optimal upper-bound corresponds to the constraint $u_{1,x}$.

The algorithm lazily explores the polyhedron by getting a model of the floating orange polyhedron.
Suppose that the first model sampled by \algref{polyReduce}, say $m_1$, has an assignment for $T$ smaller than the constant $u_{2, c}$ shown in \figref{localProjRed}.
\algref{polyReduce} calls \algref{conjBound} with $m_1$. \algref{conjBound} explores the local projection of the floating orange plane with respect to the model $m_1$. Note that on initial call to \algref{conjBound}, $T$ always has an upper bound, namely $t$. Thus \algref{conjBound} will successively project out dimensions until $T$ is unbound. In the case of \figref{localProjRed}, the dimension $y$ is locally projected out first, yielding the orange region in the $x$-$T$ plane. 
This region does have an upper-bound, $u_{1,x}$, which can simply be read off the constraint representation of the orange region. However, at this point it is unknown if there is another bound in fewer dimensions. So, \algref{conjBound} continues and locally projects out the $x$ dimension obtaining the interval $l_{1,c} \le T \le u_{2,c}$. 
There are no more dimensions to project, so \algref{conjBound} returns the conjectured upper-bound of $u_{2,c}$. Note that this is \emph{not} a true bound. That is, there is a model of the floating orange plane that is strictly larger than $u_{2,c}$.

This situation means that the while loop in \algref{polyReduce} will execute again with a new model $m_2$ that is still within the polyhedron, but is strictly larger than $u_{2,c}$. 
Thus, it will pass off this new model to \algref{conjBound}, to get new conjectured bounds. \algref{conjBound} will, again, project out the dimensions $y$ then $x$ to obtain the interval $u_{2,c}\le T$. However, in this case projecting out $y$ then $x$ gives an interval that is unbounded above. Thus, \algref{conjBound} will go back one step, when $T$ still had an upper-bound, namely $T\le u_{1,x}$. Note that the upper-bound $u_{1,x}$ is an upper-bound containing the variable $x$. Because $u_{1,x}$ is the most optimal upper-bound for $T$ using the model $m_2$, \algref{conjBound} will return $u_{1,x}$. In this case $u_{1,x}$ \emph{is} a true upper-bound. There is no value of $T$ that is strictly greater than $u_{1,x}$. For this reason, the loop in \algref{polyReduce} will terminate with $U = \{u_{2,c}, u_{1,x}\}$.
For the loop to terminate, one of these bounds must be true. So, the algorithm finishes by filtering $U$ by which ones are true bounds,
that is which $ub\in U$ have $P\models_{LRA} T\le ub$.
This check can easily be accomplished by an SMT solver. In short, such a bound must exists in $U$ because any ``upper-bound'' face of the full projection polyhedron with minimal dimensions $d_i,\dots,d_k$ is a true upper bound on $T$. Moreover, \algref{conjBound} only returns ``upper-bound'' faces of local projections that are in $d_i,\dots,d_k$ dimensions or fewer. If the upper-bound \algref{conjBound} returns is in \emph{fewer} dimensions then it's not a true bound, and another model will be sampled in \algref{polyReduce}. Since local projection finitely covers the full projection eventually a model will be sampled to produce a face of the full projection in $d_i,\dots,d_k$. For a more detailed explanation see the proof of \theoref{polyReduce}.

For this particular choice of $m_1$ and $m_2$, \algref{polyReduce} took two rounds to find a true upper-bound. However, if instead $m_2$ was selected as the first model, then the true bound would have been found and returned in only one round. For this reason, the performance of \algref{polyReduce} is heavily dependent on the models returned by the SMT solver.

\begin{theoremrep}\label{The:polyReduce}
Let $t^{*}$ be a term produced by \algref{polyReduce} for inputs $P$, $t$, and $\ll$.
Let $d_1,\dots,d_k$ be the dimensions of $P$ and $t$ sorted greatest-to-smallest with respect to $\ll$.
Let $d_j, \dots, d_k$ be the dimensions used in $t^{*}$.
Then the following are true of $t^{*}$:
\begin{enumerate}
    \item $P\models_{\textit{LRA}} t\le t^{*}$
    \item $t^{*}$ is optimal in the sense that for any other $b$ with $P\models_{\textit{LRA}} t\le b$, then $b$ is an expression over the dimensions $d_i,\dots,d_k$ with $d_j\ll d_i$.
\end{enumerate}
\end{theoremrep}
\begin{appendixproof}
To show the theorem, we first show that the full project can be used to solve the problem. Then we show how \algrefs{conjBound}{polyReduce} construct the relevant constraints of the full project in a terminating procedure. Let $d_1,\dots,d_k$ be the dimensions of $P$ and $t$ sorted greatest to smallest with respect to $\ll$. Let $P'$ be $P$ with the added constraint that $T=t$. Now let $\proj{P'}{[d_1,\dots,d_{j-1}]}$ be the most minimal polyhedron where $T$ has an upper bound. That is, $\proj{P'}{[d_1,\dots,d_{j-1}]}$ is a polyhedron where there is a constraint of the form $aT+b\ge 0$, $a<0$, $b$ is $T$ free, and $\proj{P'}{[d_1,\dots,d_{j}]}$ has no such upper-bound on $T$. In such a case, $t^{*} = -\frac{b}{a}$ is an upper-bound that satisfies the conditions of the theorem. To show this we need to show that $t^{*}$ is indeed an upper-bound, and that it is optimal.

Note that from $\proj{P'}{[d_1,\dots,d_{j-1}]}$ we have $\proj{P'}{[d_1,\dots,d_{j-1}]}\models_{LRA} T\le -\frac{b}{a}$. By properties of projection we have $\proj{P'}{[d_1,\dots,d_{j-1}]}\equiv \exists d_1,\dots,d_{j-1}. P'$. Thus, we have $P'\models_{LRA} \exists d_1,\dots,d_{j-1}. P' \equiv \proj{P'}{[d_1,\dots,d_{j-1}]}$. Furthermore, because $T$ was a fresh variable $P\equiv \exists T. (P \land T=t)$. Thus, we have the following chain of entailment:
\begin{align*}
     P\equiv &\exists T. [(P \land T = t) \land (T=t)] \models_{LRA} \exists T. [(\exists d_1,\dots,d_{j-1}. P') \land (T = t)] \equiv\\ &\exists T. [\proj{P'}{[d_1,\dots,d_{j-1}]} \land (T=t)] \models_{LRA} 
     \exists T. [T\le -\frac{b}{a} \land (T=t)] \equiv t \le -\frac{b}{a}
\end{align*}

To show optimality, suppose there was a better bound $b$. In other words there is a $b$ with $\exists T.[ (\exists d_1,\dots,d_i. P) \land (T=t)]\models_{LRA} \exists T.[T\le b \land T=t]$ with $j < i$. That is, $b$ is in a lower dimensional space than $t^{*}$. From the previous entailment and the property of projection we have $\exists T.[ (\exists d_{j+1},\dots,d_i. \proj{P}{[d_1,\dots,d_j]}) \land (T=t)]\models_{LRA} \exists T.[T\le b \land T=t]$. Further we can drop the initial existential quantifiers, and say if the previous entailment holds than so does $(\proj{P}{[d_1,\dots,d_j]}) \land (T=t)\models_{LRA} T\le b \land T=t$. However this is contradicts the assumption $\proj{P}{[d_1,\dots,d_j]}$ has $T$ unbounded above. We can get a counterexample by making $T$ arbitrarily big. This shows that $t^{*}$ is optimal.

Now we must show that \algrefs{conjBound}{polyReduce} recovers the appropriate bounds from the full projection in a finite time. First, observe that the conjectured bounds returned by \algref{conjBound} cannot be higher-dimensional than $d_j,\dots,d_k$. That is, in the dimension ordering $\ll$, the conjectured bounds returned by \algref{conjBound} are at least as good as the true optimal bound $t^{*}$. For this not to be the case, we would have to have $T$ bounded in $\lproj{m}{P'}{[d_1,\dots,d_{i-1}]}$, but not in $\lproj{m}{P'}{[d_1,\dots,d_{i}]}$ for some $d_{i+1}\gg d_j$. However, this would contradict \theoref{localProjList}, i.e. we would have $\lproj{m}{P'}{[d_1,\dots,d_{i}]} \not\models_{LRA} \proj{P'}{[d_1,\dots,d_{i}]}$. 

So, in the $\ll$ order conjectured bounds are no worse than true bounds. However, conjectured bounds can be better, in the $\ll$ order, than true bounds. In such cases the loop in \algref{polyReduce} will not terminate. That is, consider the situation where has $U$ does not contain a true bound. In such case $m\models P \bigwedge_{u\in U} \not u$ will hold, and new conjectured bounds will be produced. This shows that the while loop in \algref{polyReduce} can only terminate when a true optimal bound is conjectured.

Finally we must argue \algref{polyReduce} always terminates. This amounts to showing that \algref{polyReduce} will always produce a model such that \algref{conjBound} will return a true bound. From \theoref{localProjList} we have $\{\lproj{m}{P'}{[d_1,\dots,d_{j-1}]}\mid m\models P'\}$ finitely covers $\proj{P'}{[d_1,\dots,d_{j-1}]}$. Thus, there is some set of models such that $\lproj{m}{P'}{[d_1,\dots,d_{j-1}}$ has a constraint which is a true bound and $\lproj{m}{P'}{[d_1,\dots,d_{j}}$ has $T$ unbounded. Furthermore, the loop condition in \algref{polyReduce} ensures we continue to pick models that do not lead to the same local project. Thus, eventually some model which leads to a true bound must be chosen.
\end{appendixproof}
\theoref{polyReduce} says \algref{polyReduce} solves the OSB problem for polyhedra and ``dimensional orders''. Furthermore, due to Farkas' lemma, an optimal bound can be found with respect to polyhedral cones.
\begin{corollaryrep}\label{Cor:polyConeRed}
Let $Q$ be a polyhedral cone over $\mathbb{K}^k$ ($\mathbb{Q}^k$ or $\mathbb{R}^k$). Let $\MOLE$ be an order on the dimensions of $\mathbb{K}^k$ and $t$ a vector in $\mathbb{K}^k$. \algref{polyReduce} can be used to solve the problem of finding a $t^{*}$ with the following properties:
\begin{enumerate}
    \item $t^{*} - t \in Q$
    \item $t^{*}$ is optimal in the sense that for any other $b$ with $b-t\in Q$, then $b$ is an expression over the dimensions $d_i,\dots,d_k$ with $d_j\ll d_i$.
\end{enumerate}
\end{corollaryrep}
\begin{appendixproof}
From $Q=[q_1,\dots,q_r, \mathbf{1}]$ we can construct an appropriate polyhedron $P$ where $P^{*} = Q$. That is, make a polyhedron $P=\{q_1\ge 0, \dots, q_r\ge 0\}$. Then, by combining \theoref{polyReduce} with \lemref{linComp} we obtain the desired result.
\end{appendixproof}

\subsubsection{LP Reduction}\label{Se:lpRedux} Our main approach for reducing with respect to a polyhedral cone is \algref{polyReduce}. However, an alternative method based on linear programming is also possible. The idea is based on the observation that given a polyhedral cone $Q=[q_1,\dots, q_r, 1]$ and terms $t$ and $t'$ over dimensions $d_1,\dots,d_n$ and constant dimension $d_{n+1}$, it is possible to check whether $r-t\in Q$ with an LP query. That is, let $t^c_j$, ${t'}^c_j$, and $q^c_{i,j}$ denote the coefficient of $t$, $t'$, and $q_i$ on the $j$'th dimension. Then $t'-t\in Q$ if and only if there are non-negative $\lambda_1,\dots, \lambda_{r+1}$ with ${t'}^c_j - {t}^c_j = \lambda_1 q^c_{1,j} + \dots + \lambda_r q^c_{r,j}$ for each $1\le j\le n$ and ${t'}^{c}_{n+1} - t^c_{n+1} = \lambda_1 q^c_{1,n+1} + \dots + \lambda_r q^c_{r,n+1} + \lambda_{r+1}$. This system can be used to decide if a given concrete $t'$ has the property $t'-t\in Q$; moreover, the system can also represent the space of $t'$ with this property by leaving each ${t'}^c_j$ undetermined, i.e., by considering the linear system over the variables ${t'}^c_1,\dots, {t'}^c_{n+1}, \lambda_1,\dots,\lambda_{r+1}$. Therefore, it is possible to reduce the OSB problem over a polyhedral cone (and consequently a polyhedron) to linear programming with a lexicographic objective function \cite{Isermann1982}. Without loss of generality assume the dimensions $d_1, \dots, d_n$ are ordered in terms of preference. An optimal $t^{*}$ can be found by asking whether 
${t^{*,c}}_j - {t}^c_j = \lambda_1 q^c_{1,j} + \dots + \lambda_r q^c_{r,j}$ for each $1\le j\le n$ and ${t^{*,c}}_{n+1} - t^c_{n+1} = \lambda_1 q^c_{1,n+1} + \dots + \lambda_r q^c_{r,n+1} + \lambda_{r+1}$ has a solution with ${t^{*,c}}_1 = 0, \dots, {t^{*,c}}_n =0$. If not, then we see if there is a solution with ${t^{*,c}}_1 = 0, \dots, {t^{*,c}}_{n-1} =0$, and then ${t^{*,c}}_1 = 0, \dots, {t^{*,c}}_{n-2} =0$, etc. until a solution is found.
%We experimented with this method of solving the OSB problem for polyhedral cones along with \algref{polyReduce}. 
We found in practice \algref{polyReduce} was much faster (see \sectref{experiments}).

\subsection{Cone of Polynomials}\label{Se:coneRed}
In this section, we extend the results from the previous section to the case of a cone of polynomials.
\begin{definition}
Let $p_1,\dots,p_k$, $q_1,\dots,q_r$ be polynomials. The \emph{cone of polynomials} $C$ generated by $p_1,\dots,p_k$ and $q_1,\dots,q_r$ is the set
\[
C = \langle p_1,\dots,p_k\rangle+[q_1,\dots,q_r,1] = \{p+q\mid p\in \langle p_1,\dots,p_k\rangle, q\in [q_1,\dots,q_r,1]\}.
\]
\end{definition}
At first glance there seems to be a slight mismatch in the definition. We defined polynomial ideals as consisting of polynomials, whereas polyhedral cones are defined as a collection of vectors. However, there is no issue because polynomials can be viewed as vectors in the infinite-dimensional vector space that has monomials as basis vectors.

A cone of polynomials generated by $p_1,\dots,p_k$ and $q_1,\dots,q_r$ gives a sound set of consequences for assumptions $p_1(\mathbf{x}) = 0,\dots,p_k(\mathbf{x}) = 0$ and  $q_1(\mathbf{x})\ge 0,\dots, q_r(\mathbf{x})\ge 0$.
\begin{lemmarep}\label{Lem:coneSound}(Soundness) Let $g \in C= \langle p_1,\dots,p_k\rangle + [q_1,\dots,q_r, 1]$. Then,
\[
(\bigwedge_{i=1}^k p_i(\mathbf{x}) = 0) \land (\bigwedge_{j=1}^r q_j(\mathbf{x})\ge 0) \models_{\textit{OF}} g(\mathbf{x})\ge 0.
\]
\end{lemmarep}
\begin{appendixproof}
Because $g\in C$, there exists polynomials $h_1,\dots,h_k$ and non-negative scalars $\lambda_1,\dots,\lambda_{r+1}$ such that
\[
g = h_1p_1+\dots+h_kp_k+\lambda_1q_1+\dots+\lambda_rq_r + \lambda_{r+1}.
\]
Given the one-to-one correspondence between polynomial expressions and polynomial functions in an ordered field we have
\begin{align*}
    g(\mathbf{x}) &= h_1(\mathbf{x})p_1(\mathbf{x})+\dots+h_k(\mathbf{x})p_k(\mathbf{x})+\lambda_1q_1(\mathbf{x})+\dots+\lambda_rq_r(\mathbf{x}) + \lambda_{r+1}\\
    &= h_1(\mathbf{x})0+\dots+h_k(\mathbf{x})0+\lambda_1q_1(\mathbf{x})+\dots+\lambda_rq_r(\mathbf{x}) + \lambda_{r+1}\\
   &\ge \lambda_1 0+\dots+\lambda_r0 + \lambda_{r+1}\\
   &\ge 0\\
\end{align*}
\end{appendixproof}
\lemref{coneSound} is the main reason for our interest in cones of polynomials. Furthermore, we will show that we can perform reduction on a cone of polynomials. However, we take this moment to discuss the power of this object and the issue of completeness. In the linear case, by Farkas' lemma, polyhedral cones are complete with respect to a conjunction of linear inequalities; however, there is no such analogue for the case of a cone of polynomials\footnote{In the case of real arithmetic, positivestellensatz theorems are the analogue of Farkas' lemma for a different kind of cone object. However, they exhibit computational difficulties. See \sectref{postivestull} for more discussion.}.
That is, cones of polynomials are \emph{incomplete} for non-linear arithmetic.

\begin{example}
From an empty context, we have $\models_{\textit{NRA}} (x^n)^2 \ge 0$ for any $n\in \mathbb{N}$. If cones of polynomials were complete we would need to have $(x^n)^2$ in the ``empty cone'' $\langle 0\rangle + [1]$.
\end{example}

On the other hand because of the inclusion of the ideal, a cone of polynomials does hold onto some non-linear consequences.
\begin{example} (Extension of \exref{ideal})
$x^2-2x+2 \in \langle x-1-t,y-1-t^2\rangle + [y, 1]$
\[
x^2-2x+2 = (x-1+t)(x-1-t) + (-1)(y-1-t^2) +(1)y + (0)(1).
\]
Thus, a cone of polynomials can establish the consequence $x^2-2x+2\ge 0$ from the assumptions $x-1-t=0$, $y-1-t^2=0$, $y\ge 0$.
\end{example}
The use of cones of polynomials balances expressiveness and computational feasibility.

Before we give the method for reducing a polynomial with respect to a cone of polynomials, we need to introduce the idea of a \emph{reduced} cone of polynomials. The only difference is that in the case of reduced cone we require the $q_1,\dots,q_r$ polynomials to be reduced with respect to a Gr\"obner basis for the ideal $\langle p_1,\dots,p_k\rangle$.
\begin{definition}
Let $C=\langle p_1,\dots,p_k\rangle + [q_1,\dots,q_r,1]$ be a cone of polynomials and $\MOLE$ a monomial order. $C$ is \emph{reduced} with respect to $\MOLE$ if $p_1,\dots,p_k$ is a Gr\"obner basis for $\langle p_1,\dots,p_k\rangle$ and for every $q_i$ we have that no monomial of $q_i$ is divisible by any of $\LM(p_1),\dots,\LM(p_k)$.
\end{definition}

\begin{theoremrep}
Let $C = \langle p_1,\dots,p_k\rangle + [q_1,\dots,q_r,1]$ be a cone of polynomials. Let $B=\{g_1,\dots,g_s\}$ be a Gr\"obner basis for $\langle p_1,\dots,p_k\rangle$ and let $t_i = \textbf{red}_B(q_i)$ for each $q_i$. Then $C' = \langle g_1,\dots,g_s\rangle + [t_1,\dots,t_r,1]$ is a reduced cone with $C = C'$.
\end{theoremrep}
\begin{appendixproof}
From \theoref{grobnerBasis} we have that $C'$ is reduced, so all we need to do is show $C = C'$. First, because $g_1,\dots,g_s$ is a Gr\"obner basis we have $\langle g_1,\dots,g_s\rangle = \langle p_1,\dots,p_k\rangle$. So we need to show $C = \langle g_1,\dots,g_s\rangle + [q_1,\dots,q_r,1]$ is equal to $C'=\langle g_1,\dots,g_s\rangle + [t_1,\dots,t_r,1]$. Let $g\in C$. Then 
\begin{align*}
    g = h_1g_1+\dots + h_sg_s + \lambda_1 q_1 + \dots + \lambda_r q_r + \lambda_{r+1}
\end{align*}
for polynomials $h_1,\dots,h_s$ and non-negative scalars $\lambda_1,\dots,\lambda_{r+1}$. By construction of each $t_i$ we have $q_i = f_i + t_i$ for $f_i \in \langle g_1,\dots,g_s\rangle$.
\begin{align*}
    g =        h_1g_1+\dots + h_sg_s + \lambda_1 (&h^1_1g_1 + \dots + h^1_sg_s + t_1) + \dots +\\
    \lambda_r (&h^r_1g_1 + \dots + h^r_sg_s + t_r) +\lambda_{r+1}\\
\end{align*}
The above can be reorganized as
\begin{align*}
    g = &(h_1 + \lambda_1h^1_1 + \dots + \lambda_rh^r_1)g_1+ \dots +\\
        &(h_s + \lambda_1h^1_s + \dots + \lambda_rh^r_s)g_s+\\
        &\lambda_1t_1 + \dots \lambda_rt_r + \lambda_{r+1}
\end{align*}
Each $h_i+\lambda_1h^1_i + \dots + \lambda_rh^r_i$ is a polynomial so $g\in C'$.

For the other direction of inclusion the argument is symmetrical. Suppose $g\in C'$. Then 
\begin{align*}
    g = h_1g_1+\dots + h_sg_s + \lambda_1 t_1 + \dots + \lambda_r t_r + \lambda_{r+1}
\end{align*}
for polynomials $h_1,\dots,h_s$ and non-negative scalars $\lambda_1,\dots,\lambda_{r+1}$. By construction of each $t_i$ we have $t_i = q_i - f_i$ for $f_i \in \langle g_1,\dots,g_s\rangle$.
\begin{align*}
    g =        h_1g_1+\dots + h_sg_s + &\lambda_1 (q_1 - (h^1_1g_1 + \dots + h^1_sg_s)) + \dots +\\
                                       &\lambda_r (q_r - (h^r_1g_1 + \dots + h^r_sg_s)) +\lambda_{r+1}\\
\end{align*}
The above can be reorganized as
\begin{align*}
    g = &(h_1 - \lambda_1h^1_1 - \dots - \lambda_rh^r_1)g_1+ \dots +\\
        &(h_s - \lambda_1h^1_s - \dots - \lambda_rh^r_s)g_s+\\
        &\lambda_1q_1 + \dots \lambda_rq_r + \lambda_{r+1}
\end{align*}
Each $h_i-\lambda_1h^1_i - \dots - \lambda_rh^r_i$ is a polynomial so $g\in C$.
\end{appendixproof}

\subsubsection{Reduction} 
With the notion of a reduced cone in hand, we immediately arrive at a method to reduce a polynomial $t$ by a cone $C$ with respect to a monomial order $\MOLE$.
All we need to do is reduce $t$ by the equality part of $C$, i.e., the Gr\"obner basis, and then reduce the result by the polyhedral-cone part, using the method from \sectref{polyhedronRed}.
More explicitly, given an arbitrary monomial order $\MOLE$, cone of polynomials $C$, and polynomial $t$, we can reduce $t$ by $C$, obtaining $t^{*}$, using the following steps:
\begin{enumerate}
  \item
    From $C=\langle p_1,\dots,p_k\rangle + [q_1,\dots,q_r,1]$, compute a Gr\"obner basis $B=\{g_1,\dots,g_s\}$ for $\langle p_1,\dots,p_k\rangle$, and for each $q_i$ compute $t_i = \textbf{red}_B(q_i)$. From $B$ and $\{t_1,\dots, t_r\}$ construct the reduced cone $C' = \langle g_1,\dots,g_s\rangle + [t_1,\dots,t_r,1]$.
  \item
    ``Equality reduce'' $t$ by the Gr\"obner basis $B$.
    That is, let $t'=\textbf{red}_B(t)$.
  \item
    ``Inequality reduce'' $t'$ with \algref{polyReduce} to obtain the result $t^{*}$.
    That is, treat the polynomials of $\{t,t_1,\dots,t_r\}$ as vectors in the finite-dimensional subspace of $\mathbb{K}[\mathbf{X}]$ spanned by the monomials present in $\{t,t_1,\dots,t_r\}$, and run \algref{polyReduce}.
\end{enumerate}
\begin{definition}\label{De:coneRed}
We denote the process of reducing a polynomial $t$ by a cone of polynomials $C$ with respect to a monomial order $\MOLE$, $\textbf{cred}_C(t)$.
\end{definition}

\begin{theoremrep}\label{The:conePRed}
Let $t$ be a polynomial, $C$ a cone of polynomials, and $\MOLE$ a monomial order. Let $t^{*}= \textbf{cred}_C(t)$. $t^{*}$ has the following properties:
\begin{enumerate}
    \item $t^{*}-t\in C$
    \item $t^{*}$ is optimal in the sense that for any other $b$ with $b-t\in C$, then $\LM(b)\MOGE\LM(t^{*})$.
\end{enumerate}
\end{theoremrep}
\begin{appendixproof}
Let $C' = \langle g_1,\dots,g_s\rangle + [t_1,\dots,t_r,1]$ be a reduced cone equal to $C$. Let $t'=\textbf{red}_B(t)$. Then $t = h_1g_1+\dots h_sg_s + t'$. Let $t''$ be the result of \algref{polyReduce} on $t'$. By \corref{polyConeRed} $t'' - t'\in [t_1,\dots,t_r,1]$, so 
\begin{align*}
    t''-t' &= \lambda_1t_1+\dots+\lambda_rt_r + \lambda_{r+1}\\
    t'' - (t - (h_1g_1 + \dots + h_sg_s)) &= \lambda_1t_1+\dots+\lambda_rt_r + \lambda_{r+1}\\
    t'' - t &= (-h_1)g_1 + \dots + (-h_s)g_s + \lambda_1t_1+\dots+\lambda_rt_r + \lambda_{r+1}
\end{align*}
for non-negative $\lambda_1,\dots,\lambda_{r+1}$. The last line is the witness for $t''-t\in C' = C$.

To show optimality consider some arbitrary $b$ with $b-t\in C'$. This means
\begin{align*}
    b-t &= h'_1g_1 + \dots + h'_sg_s + \lambda'_1t_1+\dots+\lambda'_rt_r + \lambda'_{r+1}\\
    b &= t + h'_1g_1 + \dots + h'_sg_s + \lambda'_1t_1+\dots+\lambda'_rt_r + \lambda'_{r+1}
\end{align*}
Consider $\textbf{red}_B(b)=b'$. By \theoref{grobnerBasis} $b'$ has the unique property that $b = g+b'$ for $g\in \langle g_1,\dots,g_s\rangle$ and $b'$ has no term divisible by $\LM(g_1),\dots,\LM(g_s)$. Consider the term $t' + \lambda'_1t_1+\dots \lambda'_rt_r + \lambda'_{r+1}$. Because $t' = \textbf{red}_B(t)$, $t'$ has no term divisible by $\LM(g_1),\dots,\LM(g_s)$. Also because $C'$ is a reduced cone no $\lambda_it_i$ is divisible by $\LM(g_1),\dots,\LM(g_s)$ either. Finally, because $t = h_1g_1+\dots+h_sg_s +t'$ we have
\[
b = (h_1g_1+\dots+h_sg_s + h'_1g_1+\dots +h'_sg_s)+(t' + \lambda'_1t_1+\dots \lambda'_rt_r + \lambda'_{r+1})
\]
This shows that $\textbf{red}_B(b) = t' + \lambda'_1t_1+\dots \lambda'_rt_r + \lambda'_{r+1}$. So $\textbf{red}_B(b) - t' \in [t_1,\dots,t_r,1]$. However, by \corref{polyConeRed} $t''$ is the minimal term with this property. Thus, $t''$ must be an expression in fewer dimensions than $\textbf{red}_B(b)$. In other words, by interpreting $t''$ and $\textbf{red}_B(b)$ as polynomials $\LM(t'')\MOLE \LM(\textbf{red}_B(b)) \MOLE \LM(b)$. Thus, for arbitrary $b$ in $C'$, $\LM(t'')\MOLE \LM(b)$.
\end{appendixproof}
\begin{example}
Consider the example from \sectref{overview}.
In that example, we said that we had the equations $x-u_4$ and $y-u_5$ in the basis of the ideal, and the inequalities $vbu_1-u_4\ge 0$, $va2u_2+vbu_2-vbu_1+vu_2\ge 0$, and $u_5-va2u_2-vbu_2+1\ge 0$ in the basis of the polyhedral cone.
The ideal and polyhedral cone created for this example has many more equations and inequalities, but these are the ones that are relevant for reduction. 
Using the new terminology, in \sectref{overview} we reduced $t = x-y$ by the reduced cone
\[
C = \langle x-u_4,y-u_5, \dots,\rangle + [vbu_1-u_4,va2u_2+vbu_2-vbu_1+vu_2, u_5-va2u_2-vbu_2+1,\dots, 1].
\]
To reduce $x-y$ by $C$ we first equality reduce $x-y$ by the ideal, and obtain $u_4-u_5$.
Then, to reduce $u_4-u_5$ by the polyhedral cone we treat each unique monomial as a separate dimension, and run \algref{polyReduce}.
For example, we might create the map
\[
\{d_1\mapsto u_4, d_2\mapsto u_5, d_3\mapsto bvu_1, d_4\mapsto va2u_2, d_5\mapsto bvu_2, d_6\mapsto bvu_1, d_7\mapsto vu_2, \dots\}.
\]
We then reduce $d_1-d_2$ by the polyhedron
$
P \defeq \{d_3-d_1\ge 0, d_4+d_5-d_3 + d_7\ge 0, d_2-d_4-d_5+1\ge 0, \dots\}
$
and get the result $d_7+1$, or equivalently, $vu_2+1$.
By \theoref{conePRed}, $vu_2 + 1 - (x-y) \in C$, and $vu_2 + 1$ is optimal.
Furthermore, by \lemref{coneSound}, these equalities and inequalities entail $x-y\le vu_2+1$.
\end{example}

\paragraph{Optimality} \theoref{conePRed} gives optimality with respect to a monomial order $\MOLE$ which is a total order on monomials. However, when extending $\MOLE$ to polynomials, the comparison becomes a \emph{pre-order}. For example $x+y$, and $x$ have the same leading monomial if $x\MOGE y$. Furthermore, coefficients are not compared in the monomial order (for example, $5x$ and $2x$ are equivalent in the monomial order). For this reason, there can be multiple distinct optimal terms that satisfy the conditions of \theoref{conePRed}. The reduction method is not guaranteed to return all of the optimal terms. \theoref{conePRed} guarantees that the reduction will return one of the optimal bounds.

%% file: figures/localprojfig1.tex
\tdplotsetmaincoords{65}{110}
\begin{tikzpicture}
	[scale=3,
		tdplot_main_coords,
		axis/.style={->,black,thick},
		face/.style={opacity=0.5}]

	%standard tikz coordinate definition using x, y, z coords
	\coordinate (O) at (0,0,0);

	%draw axes
	\draw[axis] (0,0,0) -- (1,0,0) node[anchor=north east]{$x$};
	\draw[axis] (0,0,0) -- (0,1,0) node[anchor=north west]{$y$};
	\draw[axis] (0,0,0) -- (0,0,1) node[anchor=south]{$z$};

  \tdplotdefinepoints(0,0,0)(0.5,0.65,0.5)(0.35,0.5,0.5)
    \coordinate (0) at (0.5,0.5,0.5);
    \coordinate (a) at (\tdplotax,\tdplotay,\tdplotaz);
    %\node at (a) [below = 1mm of a] {a};
    \coordinate (at) at (\tdplotax, \tdplotay, 0.75);
    \coordinate (ap) at (\tdplotax, \tdplotay, 0);
    \coordinate (b) at (\tdplotbx,\tdplotby,\tdplotbz);
    %\node at (b) [below = 1mm of b] {b};
    \coordinate (bt) at (\tdplotbx, \tdplotby, 0.75);
    \coordinate (bp) at (\tdplotbx, \tdplotby, 0);
    
    \tdplotdefinepoints(0,0,0)(0.65,0.5,0.5)(0.5,0.35,0.5)
    \coordinate (c) at (\tdplotax,\tdplotay,\tdplotaz);
    %\node at (c) [below = 1mm of c] {c};
    \coordinate (ct) at (\tdplotax, \tdplotay, 0.75);
    \coordinate (cp) at (\tdplotax, \tdplotay, 0);
    \coordinate (d) at (\tdplotbx,\tdplotby,\tdplotbz);
    %\node at (d) [below = 1mm of d] {d};
    \coordinate (dt) at (\tdplotbx, \tdplotby, 0.75);
    \coordinate (dp) at (\tdplotbx, \tdplotby, 0);
    
    \tdplotdefinepoints(0,0,0)(0.75,0.75,0.75)(0.25,0.75,0.75)
	\coordinate (1) at (\tdplotax,\tdplotay,\tdplotaz);
	%\node at (1) [right = 1mm of 1] {1};
	\coordinate (1p) at (\tdplotax,\tdplotay,0);
	\coordinate (2) at (\tdplotbx,\tdplotby,\tdplotbz);
	%\node at (2) [above = 1mm of 2] {2};
	\coordinate (2p) at (\tdplotbx,\tdplotby,0);
	
 \draw[dashed] (a) -- (b)  -- (d);
 \draw[dashed] (b) -- (2);

 \path[fill=blue!60,opacity=0.8] (2) -- (bt) -- (at) -- cycle;
\path[fill=blue!60,opacity=0.8] (b) -- (a) -- (at) -- (bt) -- cycle;
\path[fill=blue!60] (ap) -- (bp) -- (2p) -- cycle;

 \path[fill=gray!60,opacity=0.8] (c) -- (d) -- (dt) -- (ct) -- cycle;
  \path[fill=gray!60,opacity=0.8] (ct) -- (dt) -- (bt) -- (at)--cycle;

 \path[fill=gray!60] (cp) -- (dp) -- (bp) -- (ap)--cycle;

  \path[fill=olive!60,opacity=0.8] (a) -- (2) -- (1) -- cycle;
 \path[fill=olive!60] (ap) -- (2p) -- (1p) -- cycle;
 \path[fill=olive!60,opacity=0.8] (1) -- (at) -- (2) -- cycle;
 \path[fill=olive!60,opacity=0.8] (a) -- (at) -- (1) -- cycle;
  
   \path[fill=red!60,opacity=0.8] (1) -- (c) -- (ct) -- cycle;
 \path[fill=red!60,opacity=0.8] (c) -- (a) -- (1) -- cycle;
 \path[dotted,fill=red!60] (cp) -- (ap) -- (1p) -- cycle;
  \path[dotted,fill=red!60,opacity=0.8] (1) -- (ct) -- (at) -- cycle;

  \draw[] (d) -- (c) -- (a);
 \draw[] (d) -- (dt);
 \draw[] (c) -- (1) -- (a);
 \draw[] (c) -- (ct);
 \draw[] (a) -- (2);
 \draw[] (2) -- (bt) -- (dt) -- (ct) -- (1) -- cycle;

\end{tikzpicture}

%% file: figures/localprojfig2.tex
\tdplotsetmaincoords{65}{110}
\begin{tikzpicture}
	[scale=3,
		tdplot_main_coords,
		axis/.style={->,black,thick},
		face/.style={opacity=0.5}]

	%standard tikz coordinate definition using x, y, z coords
	\coordinate (O) at (0,0,0);

	%draw axes
	\draw[axis] (0,0,0) -- (1,0,0) node[anchor=north east]{$x$};
 	\draw[axis] (0,0,0) -- (0,1,0) node[anchor=north west]{$y$};
	\draw[axis] (0,0,0) -- (0,0,1) node[anchor=south]{$z$};

  \tdplotdefinepoints(0,0,0)(0.5,0.65,0.5)(0.35,0.5,0.5)
    \coordinate (0) at (0.5,0.5,0.5);
    \coordinate (a) at (\tdplotax,\tdplotay,\tdplotaz);
    %\node at (a) [below = 1mm of a] {a};
    \coordinate (at) at (\tdplotax, \tdplotay, 0.75);
    \coordinate (atp) at (\tdplotax, 0, 0.75);
    \coordinate (ap) at (\tdplotax, 0, \tdplotaz);
    \coordinate (b) at (\tdplotbx,\tdplotby,\tdplotbz);
    %\node at (b) [below = 1mm of b] {b};
    \coordinate (bt) at (\tdplotbx, \tdplotby, 0.75);
    \coordinate (btp) at (\tdplotbx, 0,0.75);
    \coordinate (bp) at (\tdplotbx, 0, \tdplotbz);
    \coordinate (bpypzgr) at (\tdplotbx, 0, 0.04);
    \coordinate (bpypzb) at (\tdplotbx, 0, 0.04);
    
    \tdplotdefinepoints(0,0,0)(0.65,0.5,0.5)(0.5,0.35,0.5)
    \coordinate (c) at (\tdplotax,\tdplotay,\tdplotaz);
    %\node at (c) [below = 1mm of c] {c};
    \coordinate (ct) at (\tdplotax, \tdplotay, 0.75);
    \coordinate (ctp) at (\tdplotax, 0, 0.75);
    \coordinate (cp) at (\tdplotax, 0,\tdplotaz);
    \coordinate (cpypzr) at (\tdplotax, 0,0.04);
    \coordinate (cpypzg) at (\tdplotax, 0,0.04);
    \coordinate (d) at (\tdplotbx,\tdplotby,\tdplotbz);
    %\node at (d) [below = 1mm of d] {d};
    \coordinate (dt) at (\tdplotbx, \tdplotby, 0.75);
    \coordinate (dp) at (\tdplotbx, 0,\tdplotbz);
    \coordinate (dpypz) at (\tdplotbx, 0,0.04);
    
    \tdplotdefinepoints(0,0,0)(0.75,0.75,0.75)(0.25,0.75,0.75)
	\coordinate (1) at (\tdplotax,\tdplotay,\tdplotaz);
	%\node at (1) [right = 1mm of 1] {1};
	\coordinate (1p) at (\tdplotax,0,\tdplotaz);
	\coordinate (1pypzr) at (\tdplotax,0,0.04);
	\coordinate (2) at (\tdplotbx,\tdplotby,\tdplotbz);
	%\node at (2) [above = 1mm of 2] {2};
	\coordinate (2p) at (\tdplotbx,0,\tdplotbz);
	\coordinate (2pypzb) at (\tdplotbx,0,0.04);
	
 \draw[dashed] (a) -- (b)  -- (d);
 \draw[dashed] (b) -- (2);

 \coordinate (ct21int) at (0.65,0.75,0.75);
 \coordinate (ct21inta1int) at (0.65, 0.71,0.65);
 \coordinate (dt21int) at (0.5,0.75,0.75);
 \coordinate (bt21int) at (0.35,0.75,0.75);
 \coordinate (bt21inta2int) at (0.35,0.71,0.65);

 \path[fill=blue!60,opacity=0.8] (bt) -- (2) -- (bt21int) -- cycle;
 \path[fill=blue!60,opacity=0.8] (bt) -- (b) -- (bt21inta2int) -- (bt21int) -- cycle;
 \path[fill=blue!60,opacity=0.8] (bt21inta2int) -- (bt21int) -- (2) -- cycle;
 \path[fill=blue!60,opacity=0.8] (b) -- (bt21inta2int) -- (2) -- cycle;
 
 \path[fill=gray!60,opacity=0.8] (dt) -- (dt21int) -- (bt21int) -- (bt) -- cycle;
  \path[fill=gray!60,opacity=0.8] (d) -- (a) -- (dt21int) -- (dt) -- cycle;
\path[fill=gray!60,opacity=0.8] (dt21int) -- (bt21int) -- (bt21inta2int) -- (a) -- cycle;
\path[fill=gray!60,opacity=0.8] (a) -- (b) -- (bt21inta2int) -- cycle;
 
  \path[fill=green!60,opacity=0.6] (c) -- (a) -- (ct21inta1int) -- cycle;
  \path[fill=green!60,opacity=0.6] (a) -- (ct21inta1int) -- (ct21int) -- (dt21int) -- cycle;
  \path[fill=green!60,opacity=0.6] (dt) -- (ct) -- (ct21int) -- (dt21int) -- cycle;
  \path[fill=green!60,opacity=0.6] (d) -- (c) -- (ct) -- (dt) -- cycle;
 
 \path[fill=red!60,opacity=0.6] (ct) -- (ct21int) --(1) -- cycle;
 \path[fill=red!60,opacity=0.6] (ct21int) -- (ct21inta1int) -- (1) -- cycle;
 \path[fill=red!60,opacity=0.6] (ct21inta1int) -- (c) -- (1) -- cycle;
 \path[fill=red!60,opacity=0.6] (c) -- (ct) -- (1) -- cycle;

 \path[fill=blue!60] (2p) -- (bp) -- (btp) -- cycle;
 \path[fill=gray!60] (bp) -- (btp) -- (atp) -- (ap) -- cycle;
 \path[fill=green!60] (ap) -- (cp) -- (ctp) -- (atp) -- cycle;
 \path[fill=red!60] (cp) -- (ctp) -- (1p) -- cycle;
 
 %\draw[dotted] (dt) -- (dt21int);
 %\draw[dotted] (ct) -- (ct21int);
 %\draw[dotted] (c) -- (ct21int);
 %\draw[dotted] (bt) -- (bt21int);
 %\draw[dotted] (ct21int) -- (ct21inta1int);
 %\draw[dotted] (ct21inta1int) -- (c);
 %\draw[dotted] (a) -- (dt21int);
 
 \draw[] (d) -- (c) -- (a);
 \draw[] (d) -- (dt);
 \draw[] (c) -- (1) -- (a);
 \draw[] (c) -- (ct);
 \draw[] (a) -- (2);
 \draw[] (2) -- (bt) -- (dt) -- (ct) -- (1) -- cycle;
 
 \draw[red] (1pypzr) -- (cpypzr);
 \draw[green] (cpypzg) -- (dpypz);
 \draw[gray] (dpypz) -- (bpypzgr);
  \draw[blue] (2pypzb) -- (bpypzb);
\end{tikzpicture}

%% file: figures/localprojfig3.tex
\tdplotsetmaincoords{65}{110}
\begin{tikzpicture}
	[scale=3,
		tdplot_main_coords,
		axis/.style={->,black,thick},
		face/.style={opacity=0.5}]

	%standard tikz coordinate definition using x, y, z coords
	\coordinate (O) at (0,0,0);

	%draw axes
	\draw[axis] (0,0,0) -- (1,0,0) node[anchor=north east]{$x$};
	\draw[axis] (0,0,0) -- (0,1,0) node[anchor=north west]{$y$};
	\draw[axis] (0,0,0) -- (0,0,1) node[anchor=south]{$z$};

  \tdplotdefinepoints(0,0,0)(0.5,0.65,0.5)(0.35,0.5,0.5)
    \coordinate (0) at (0.5,0.5,0.5);
    \coordinate (a) at (\tdplotax,\tdplotay,\tdplotaz);
    %\node at (a) [below = 1mm of a] {a};
    \coordinate (at) at (\tdplotax, \tdplotay, 0.75);
    \coordinate (ap) at (\tdplotax, \tdplotay, 0);
    \coordinate (b) at (\tdplotbx,\tdplotby,\tdplotbz);
    %\node at (b) [below = 1mm of b] {b};
    \coordinate (bt) at (\tdplotbx, \tdplotby, 0.75);
    \coordinate (bp) at (\tdplotbx, \tdplotby, 0);
    
    \tdplotdefinepoints(0,0,0)(0.65,0.5,0.5)(0.5,0.35,0.5)
    \coordinate (c) at (\tdplotax,\tdplotay,\tdplotaz);
    %\node at (c) [below = 1mm of c] {c};
    \coordinate (ct) at (\tdplotax, \tdplotay, 0.75);
    \coordinate (cp) at (\tdplotax, \tdplotay, 0);
    \coordinate (d) at (\tdplotbx,\tdplotby,\tdplotbz);
    %\node at (d) [below = 1mm of d] {d};
    \coordinate (dt) at (\tdplotbx, \tdplotby, 0.75);
    \coordinate (dp) at (\tdplotbx, \tdplotby, 0);
    
    \tdplotdefinepoints(0,0,0)(0.75,0.75,0.75)(0.25,0.75,0.75)
	\coordinate (1) at (\tdplotax,\tdplotay,\tdplotaz);
	%\node at (1) [right = 1mm of 1] {1};
	\coordinate (1p) at (\tdplotax,\tdplotay,0);
	\coordinate (2) at (\tdplotbx,\tdplotby,\tdplotbz);
	%\node at (2) [above = 1mm of 2] {2};
	\coordinate (2p) at (\tdplotbx,\tdplotby,0);
	
 \draw[dashed] (a) -- (b)  -- (d);
 \draw[dashed] (b) -- (2);

  \coordinate (ct21int) at (0.65,0.75,0.75);
 \coordinate (ct21inta1int) at (0.65, 0.71,0.65);
 \coordinate (dt21int) at (0.5,0.75,0.75);
 \coordinate (bt21int) at (0.35,0.75,0.75);
 \coordinate (bt21inta2int) at (0.35,0.71,0.65);
 \coordinate (bt2atint) at (0.35,0.71,0.75);
  \coordinate (ct1atint) at (0.65,0.71,0.75);
 
 \path[fill=blue,opacity=0.8] (2) -- (bt2atint) -- (bt) -- cycle;
 
 \path[fill=blue!60,opacity=0.8] (at) -- (bt2atint) -- (bt) -- cycle;
\path[fill=blue!60,opacity=0.8] (b) -- (a) -- (at) -- (bt) -- cycle;

 \path[fill=gray,opacity=0.8] (dt) -- (at) -- (bt) -- cycle;
 \path[fill=gray,opacity=0.8] (d) -- (dt) -- (at) -- (a) -- cycle;
 
 \path[fill=gray!60,opacity=0.8] (dt) -- (at) -- (ct) -- cycle;
 \path[fill=gray!60,opacity=0.8] (c) -- (d) -- (dt) -- (ct) -- cycle;

\path[fill=olive,opacity=0.7] (2) -- (at) -- (dt21int) -- cycle;
\path[fill=olive,opacity=0.7] (2) -- (dt21int) -- (a) -- cycle;
\path[fill=olive!60,opacity=0.7] (a) -- (dt21int) -- (1) -- cycle;
\path[fill=olive!60,opacity=0.7] (1) -- (dt21int) -- (at) -- cycle;

 %\path[fill=green!60,opacity=0.8] (a) -- (2) -- (1) -- cycle;
 %\draw[dotted,fill=green!60] (ap) -- (2p) -- (1p) -- cycle;
 %\draw[dotted,fill=green!60,opacity=0.8] (1) -- (at) -- (2) -- cycle;
 %\draw[dotted,fill=green!60,opacity=0.8] (a) -- (at) -- (1) -- cycle;
  
 \path[fill=red,opacity=0.6] (ct) -- (at) -- (ct1atint) -- cycle;
 \path[fill=red,opacity=0.6] (c) -- (ct21inta1int) -- (a) -- cycle;
 \path[fill=red!60,opacity=0.6] (ct) -- (1) -- (ct1atint) -- cycle;
 \path[fill=red!60,opacity=0.6] (c) -- (ct) -- (1) -- cycle;
 \path[fill=red!60,opacity=0.6] (c) -- (ct21inta1int) -- (1) -- cycle;
 
 %\path[fill=red!60,opacity=0.6] (c) -- (a) -- (1) -- cycle;

  %\path[fill=red!60,opacity=0.8] (1) -- (c) -- (ct) -- cycle;
%\path[fill=red!60,opacity=0.8] (c) -- (a) -- (1) -- cycle;
% \path[dotted,fill=red!60] (cp) -- (ap) -- (1p) -- cycle;
%  \path[dotted,fill=red!60,opacity=0.8] (1) -- (ct) -- (at) -- cycle;
 
 \path[fill=gray] (dp) -- (ap) -- (bp) --cycle;
  \path[fill=gray!60] (dp) -- (ap) -- (cp) --cycle;
  
  \path[fill=blue!60] (ap) -- (bp) -- (0.35,0.71,0) -- cycle;
  \path[fill=blue] (bp) -- (0.35,0.71,0) -- (2p) -- cycle;
  
  \path[fill=olive] (2p) -- (0.5,0.75,0) -- (ap) -- cycle;
  \path[fill=olive!60] (1p) -- (0.5,0.75,0) -- (ap) -- cycle;
  
  \path[fill=red] (cp) -- (ap) -- (0.65, 0.71,0) -- cycle;
  \path[fill=red!60] (cp) -- (1p) -- (0.65, 0.71,0) -- cycle;

  \draw[] (d) -- (c) -- (a);
 \draw[] (d) -- (dt);
 \draw[] (c) -- (1) -- (a);
 \draw[] (c) -- (ct);
 \draw[] (a) -- (2);
 \draw[] (2) -- (bt) -- (dt) -- (ct) -- (1) -- cycle;
 
 \draw[gray] (0.35,0.03,0) -- (0.5,0.03,0);
 \draw[gray!60] (0.5,0.02,0) -- (0.65,0.02,0);
 
 \draw[blue] (0.25,0.05,0) -- (0.35,0.05,0);
 \draw[blue!60] (0.35,0.06,0) -- (0.5,0.06,0);
 
 \draw[red] (0.5,0.05,0) -- (0.65,0.05,0);
 \draw[red!60] (0.65,0.06,0) -- (0.75,0.06,0);
 
 \draw[olive] (0.25,0.1,0) -- (0.5,0.1,0);
 \draw[olive!60] (0.5,0.09,0) -- (0.75,0.09,0);
 
\end{tikzpicture}

%% file: figures/localprojfig4.tex
\tdplotsetmaincoords{65}{120}
\begin{tikzpicture}
	[scale=4,
		tdplot_main_coords,
		axis/.style={->,black,thick},
		face/.style={opacity=0.5}]

	%standard tikz coordinate definition using x, y, z coords
	\coordinate (O) at (0,0,0);

	%draw axes
	\draw[axis] (0,0,0) -- (1,0,0) node[anchor=north east]{$x$};
	\draw[axis] (0,0,0) -- (0,1,0) node[anchor=north west]{$y$};
	\draw[axis] (0,0,0) -- (0,0,1) node[anchor=south]{$T$};

  % 0.5(x-0.2) + (y-0.2) = D
  \tdplotdefinepoints(0.3,0.5,0.03469)(0.15,0.65,0.2)(0.5,1,0.8)
  \coordinate (1) at (\tdplotvertexx, \tdplotvertexy, \tdplotvertexz);
  \coordinate (1yp) at (\tdplotvertexx, 0, \tdplotvertexz);
  \coordinate (1xp) at (\tdplotvertexx, \tdplotvertexy, 0);
  \coordinate (1ypxp) at (0, 0, \tdplotvertexz);
  \coordinate (1t) at (\tdplotvertexx, \tdplotvertexy, 1);
  \coordinate (1b) at (\tdplotvertexx, \tdplotvertexy, -0.2);
  %\node[] at (1) [below =1mm of 1] {1};
  
  \coordinate (2) at (\tdplotax, \tdplotay, \tdplotaz);
  \coordinate (2yp) at (\tdplotax, 0, \tdplotaz);
  \coordinate (2xp) at (\tdplotax, \tdplotay, 0);
  \coordinate (2ypxp) at (0, 0, \tdplotaz);
  \coordinate (2t) at (\tdplotax, \tdplotay, 1);
  \coordinate (2b) at (\tdplotax, \tdplotay, -0.2);
%\node[] at (2) [below =1mm of 2] {2};

  \coordinate (3) at (\tdplotbx, \tdplotby, \tdplotbz);
  \coordinate (3yp) at (\tdplotbx, 0, \tdplotbz);
  \coordinate (3xp) at (\tdplotbx, \tdplotby, 0);
  \coordinate (3ypxp) at (0, 0, \tdplotbz);
  \coordinate (3t) at (\tdplotbx, \tdplotby, 1);
  \coordinate (3b) at (\tdplotbx, \tdplotby, -0.2);
  %\node[] at (3) [above =1mm of 3] {3};
  
  \tdplotdefinepoints(0.9,0.7,0.5)(0,0,0)(0,0,0)
  \coordinate (4) at (\tdplotvertexx, \tdplotvertexy, \tdplotvertexz);
  \coordinate (4yp) at (\tdplotvertexx, 0, \tdplotvertexz);
  \coordinate (4xp) at (\tdplotvertexx, \tdplotvertexy, 0);
  \coordinate (4ypxp) at (0, 0, \tdplotvertexz);
  \coordinate (4t) at (\tdplotvertexx, \tdplotvertexy, 1);
  \coordinate (4b) at (\tdplotvertexx, \tdplotvertexy, -0.2);
  %\node[] at (4) [above =1mm of 4] {4};

 \path[fill=blue!40, opacity=0.3] (1xp) -- (2xp) -- (2t) -- (1t) -- cycle;
  \path[fill=blue!40, opacity=0.3] (1xp) -- (4xp) -- (4t) -- (1t) -- cycle;
   \path[fill=blue!40, opacity=0.3] (2xp) -- (3xp) -- (3t) -- (2t) -- cycle;
 \path[fill=blue!40, opacity=0.3] (1xp) -- (4xp) -- (4b) -- (1b) -- cycle;
 \path[fill=blue!40, opacity=0.3] (1xp) -- (2xp) -- (2b) -- (1b) -- cycle;
 \path[fill=blue!40, opacity=0.3] (2xp) -- (3xp) -- (3b) -- (2b) -- cycle;
  \draw[dotted, ->] (1xp) -- (1t);
 \draw[dotted, ->] (2xp) -- (2t);
 \draw[dotted, ->] (1xp) -- (1b);
 \draw[dotted, ->] (2xp) -- (2b);
 %\path[fill=blue!40, opacity=0.3] (3xp) -- (4xp) -- (4t) -- (3t) -- cycle;
 
  \path[top color =orange!30, bottom color = orange] (1) -- (4) -- (3) -- (2) -- cycle;
  \draw[dashed, ->] (1) -- (4);
  \draw[dashed] (1) -- (2);
  \draw[dashed, ->] (2) -- (3);
 
\draw[orange, line width = 0.6mm] (1ypxp) -- (2ypxp);
\node at (1ypxp) [right=1mm] {$l_{1,c}$};
\draw[orange, line width = 1mm,->] (2ypxp) -- (0, 0, 0.7);
\node at (2ypxp) [right=1mm] {$u_{2,c}$};
 
 \path[top color =orange!30, bottom color = orange] (1yp) -- (4yp) -- (3yp) -- (2yp) -- cycle;
 \draw[] (1yp) node[fill=white, right= 1mm] {$l_{1,x}$} -- (2yp);
 %\node[below right= 1mm of 1yp] {$l_{1,x}$}
 \draw[->] (1yp) -- node[left=1mm] {$l_{2,x}$} (4yp);
 \draw[->] (2yp) -- (3yp);
 \node at (3yp) [above=1mm] {$u_{1,x}$};
 
 \path[top color=blue!40, bottom color = blue!20] (1xp) -- (4xp) -- (3xp) -- (2xp) -- cycle;
 \draw[] (1xp) -- (2xp);
 \draw[->] (1xp) -- (4xp);
 \draw[->] (2xp) -- (3xp);

\end{tikzpicture}

%% file: saturation.tex
\section{Saturation}\label{Se:saturation}
% We suspect that many practical problems require a setting more general than cones of polynomials. That is, rather than optimizing with respect to a cone,
% a more useful capability is to be able to optimize with respect to
% an arbitrary quantifier-free \emph{non-linear} formula $\phi$, which may contain additional function symbols outside the signature of polynomial arithmetic.
% As mentioned before, solving the general non-linear problem in the case of rational arithmetic is impossible, and very expensive in the case of real arithmetic.
% Thus, we expect that any method that attempts to address this problem
% must be
% heuristic in nature.

% \yf{Do we need the previous paragraph? matches the problem statement and method outline in the overview. I didn't earlier get the feeling that the previous section solves OSB w.r.t.\ a cone of polynomials, because I understand that as a technical object in the process of solving the OSB problem from the introduction.}
In this section, we give a heuristic method for the non-linear OSB problem. The idea is that from a formula $\phi$ we extract implied polynomial equalities and polynomial inequalities, and construct a cone of polynomials from the result.
We illustrate the process using the example from \sectref{overview}.

\paragraph{Purification} The first step is to \emph{purify} all the terms in the formula. For each non-polynomial function symbol, we introduce a fresh variable, replace all occurrences of the function with the introduced symbol, and remember the assignment in a foreign-function map. We also perform this process with respect to the function arguments as well. Thus, the foreign-function map consists of assignments $u\mapsto f(p)$, where $f$ is a non-polynomial function symbol, but $p$ has also been purified and is therefore a polynomial.
The result of purification on the example from \sectref{overview} results in the following formula and foreign-function map:
\begin{align*}
    \phi' \defeq\ &x = u_4 \land y = u_5\land a2 = u_6 \land e' = e+a \land b' = b + a2 \land a,b,e,v \ge 0\\
    TM \defeq\ &\{u_1\mapsto e^{-1}, u_2 \mapsto e'^{-1}, u_3 \mapsto e^{-1}, u_4 \mapsto \floor{vbu_1}, u_5 \mapsto \floor{vb'u_2}, u_6\mapsto \floor{abu_3}\}
\end{align*}
As mentioned in \sectref{overview}, we can also consider additional axioms satisfied by the non-polynomial functions.
For example, $e\ge 0\implies \frac{1}{e}\ge 0$.
Formally, we consider these instantiated axioms as being provided by the user in the original formula;
however, for convenience, in our implementation and experiments, we use templates and instantiate these axioms automatically using the foreign-function map, and then conjoin the result onto the original formula.

\begin{wrapfigure}{R}{0.62\textwidth}
\begin{minipage}{0.61\textwidth}
\vspace{-0.8cm}
\begin{figure}[H]
\centering
    \input{figures/consat}
    \vspace{-0.7cm}
    \caption{Cone saturation overview}
    \label{Fi:coneSat}
\end{figure}
\vspace{-0.6cm}
\end{minipage}
\end{wrapfigure}
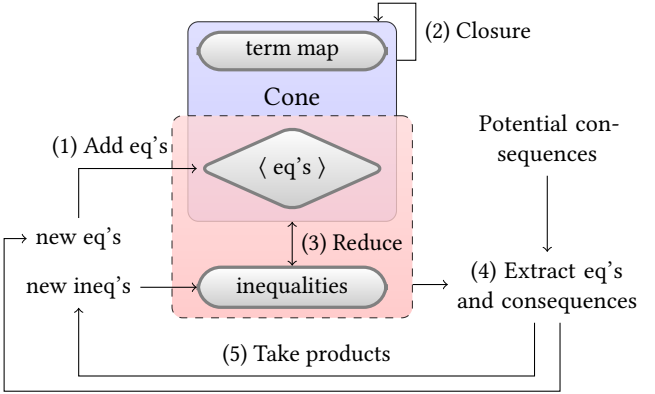
Once the formula has been purified and a function map created, the task becomes to extract an implied cone from the combination of the function map and the purified formula. We refer to this process as \emph{saturation}. Within a saturation step, two flavors of implied equalities and inequalities can be produced.
There are ones that are (linearly) implied by the formula, and there are others that are \emph{implied} by the cone, but not \emph{in} the cone.
For example, consider the cone $C_1=\langle 0\rangle +[x,1]$.
$x\in C$ corresponds to $x\ge 0$, which implies $x^2\ge 0$;
however, $x^2\notin C_1$.
If we add $x^2$ to $C_1$, we get $C_2=\langle 0\rangle + [x^2,x,1]$ with $C_1\subsetneq C_2$. 
Such a step, where we add implied equalities and inequalities to a cone that are not members of the cone, is referred to as a \emph{strengthening} step. 
\figref{coneSat} gives an overview of our saturation method. The process is iterative until no more equalities or inequalities can be added to the cone of polynomials.

\subsection{Equality Saturation}\label{Se:closure}
This subsection covers steps (1), (2), and (3)
of \figref{coneSat}.
We first assume we have some new implied equations $\{p_1=0, \dots,p_l=0\}$ and inequalities $\{q_1\ge 0, \dots, q_r\ge 0\}$, which have been produced from a yet-to-be-explained method.
We take the equations, add them to an ideal, and compute a Gr\"obner basis;
we take the inequalities and add them to a polyhedral cone.
For the example from \sectref{overview}, suppose that we are given the equations
$x = u_4$, $y = u_5$, $a2 = u_6$, $e' = e+a$, $b' = b + a2$, $eu_1=1$, $e'u_2=1$, $eu_3=1$, and no inequalities.
We add the equalities to an ideal and compute a Gr\"obner basis, which for this example would yield:
$\langle x-u_4, y-u_5, a2-u_6, e'-e-a, b'-b-a2, eu_1-1, eu_2+au_2-1, eu_3-1\rangle$.
We have now finished step (1), with 
the following cone, map, and formula with instantiated axioms:
\begin{align*}
    \phi''\defeq\ &a, b, e, v\ge 0 \land (e\ge 0\implies u_1\ge 0) \land (vbu_1\ge 0\implies u_4\ge 0)\land \dots\\
    C \defeq\ &\langle x-u_4, y-u_5, a2-u_6, e'-e-a, b'-b-a2, eu_1-1, eu_2+au_2-1, eu_3-1\rangle + [1]\\
    TM \defeq\ &\{u_1\mapsto e^{-1}, u_2 \mapsto e'^{-1}, u_3\mapsto e^{-1}, u_4 \mapsto \floor{vbu_1}, u_5 \mapsto \floor{vb'u_3}, u_6\mapsto \floor{abu_2}\}
\end{align*}

Step (2) is a strengthening step, where we perform a type of congruence-closure process on the ideal and the foreign-function map. Consider the running example.
In the foreign-function map we have $u_1\mapsto e^{-1}$ and $u_3\mapsto e^{-1}$.
By the axiom $w=w' \implies (w) = f(w')$, for any function $f$,
it is clear that $u_1=u_3$.
However, $u_1-u_3$ is \emph{not} a member of the ideal. The purpose of step (2) (Closure) is to find these equalities. 

Our closure algorithm works by considering each pair of assignments $w\mapsto f(p)$ and $w'\mapsto f'(p')$ in the foreign-function map, where $f$ and $f'$ are the same function symbol. To check if the arguments are equal we check whether $p-p'$ is a member of the ideal which by \corref{idealMem} can be done by checking if the result of reducing $p-p'$ by the Gr\"obner basis is 0. If the result is $0$ then we have that the ideal entails $p=p'$, so $f(p)=f'(p')$ and $w=w'$. In this case we add the new equality $w-w'$ (meaning $w-w'=0$) into the ideal and compute a new Gr\"obner basis. The new ideal might uncover new equalities of the map, so we have to check each pair of functions again until no new equalities are discovered.
\begin{lemmarep}
Closure terminates.
\end{lemmarep}
\begin{appendixproof}
By Hilbert's basis theorem we can only add finitely many equations to the ideal before it eventually stabilizes. Once the ideal there are only finitely many pairs of functions to check for equality, so closure will eventually terminate.
\end{appendixproof}

After equality saturation, we reduce the set of inequalities by the newly returned Gr\"obner basis to keep the cone reduced in the sense of \defref{coneRed}.
This reduction is step (3) in \figref{coneSat}.
Returning to the running example, after step (3) we have the following cone, foreign-function map,
and formula with instantiated axioms:
\begin{align*}
    \phi''\defeq\ &a, b, e, v\ge 0 \land (e\ge 0\implies u_1\ge 0) \land (vbu_1\ge 0\implies u_4\ge 0)\land \dots\\
    C \defeq\ &\langle x-u_4, y-u_5, a2-u_6, e'-e-a, b'-b-a2, eu_1-1, eu_2+au_2-1, u_3-u_1\rangle + [1]\\
    TM \defeq\ &\{u_1\mapsto e^{-1}, u_2 \mapsto e'^{-1}, u_3\mapsto e^{-1}, u_4 \mapsto \floor{vbu_1}, u_5 \mapsto \floor{vb'u_2}, u_6\mapsto \floor{abu_3}\}
\end{align*}

\subsection{Consequence Finding}\label{Se:consFinding}
In step (4), our goal is two-fold.
First, we want to make the polyhedral cone of inequalities salient in the sense of \defref{salient};
second, we want to extract inequalities implied by the formula.
For both of these goals, we generate a set of potential consequences and use a \emph{linear} SMT solver to filter the potential consequences down to a set of true consequences.

We first need to explain the relevance of making the polyhedral cone salient.
Suppose that we had the cone of polynomials $C=\langle p_1,\dots, p_k\rangle + [q_1,\dots,q_r,1]$.
The polyhedral cone $P = [q_1,\dots,q_r,1]$ represents inequalities;
i.e., $v\in P$ corresponds to $v\ge 0$.
If $P$ is not salient, then there exists a polynomial $f\in P$ and $-f\in P$, implying that we can derive $f\ge 0$ and $-f\ge 0$, so $f=0$;
however, assuming that the cone was reduced, $f \notin \langle p_1, \dots, p_k \rangle$.
Ideal reasoning is stronger than polyhedral-cone reasoning, so in this situation if we created $C'=\langle f, p_1,\dots, p_k\rangle + Q$, where $Q$ is $P$ with $f$ removed, we would have $C\subsetneq C'$.
Fortunately, we can reformulate \lemref{salient} to say that if there are no implied equations among $\{q_1,\dots,q_r\}$ then there are no implied equations in all of $P$.
Thus, we can make $P$ salient by asking if any of the equalities $q_1=0, \dots, q_r= 0$ are implied. If so, these are newly discovered equations that will be added to the ideal in step (1) on the next saturation round.

Also, in this step we want to extract other inequalites that are implied by the formula. Consider the running example. We have that $C$ implies $e\ge 0$ and from $\phi''$ we have $e\ge0 \implies u_1\ge 0$. Thus, $u_1\ge 0$, but $u_1$ is \emph{not} a member of $C$. To extract $u_1\ge 0$ as a true consequence, we generate a finite list of potential consequences by adding each atom of the negation normal form of $\phi''$ as a potential consequence.
For example, from $\phi''$ some potential consequences are $e<0$, $u_1\ge0$, $vbu_1<0$, and $u_4\ge0$.
Note that even in the linear case this method is incomplete.
That is, there are inequalities that are implied by the formula that are not present in the formula.
For example, $(x \ge 0) \land (x \le 1  \implies y\ge x)\land (1\le x\le 2 \implies y\ge -x+2) \land (x\ge 2\implies y\ge \frac{1}{2}x-1)$ entails $y\ge 0$, but $y\ge 0$ is not found in any atom of the negation normal form of the formula.\footnote{In this step we can also extract equalities. We can generate potential equalities along with inequalities by looking at the formula for equality atoms. However, if we only want to look for inequalities, saturation will still work because inequalities that are actually equalities will get upgraded in some later round of saturation.}

We collect both the potential equality consequences and the potential inequality consequences into a list, reduce them with the ideal, and then use a Houdini \cite{houdini} like algorithm using a \emph{linear} SMT solver to filter the potential consequences to a list of true consequences. We use a linear SMT solver as opposed to a non-linear one to avoid the aforementioned issues of non-linear reasoning. That is, we replace each monomial with a fresh variable before determining true consequences.
For the running example, we do not yet have any known inequalities, so we do not have any inequalities to potentially upgrade to equalities;
however, from the formula $\phi''$ we generate the potential consequences $\mathit{Cons} =\{e\ge0,b\ge 0, e\ge0, v\ge 0, e<0, u_1\ge 0, vbu_1 < 0, u_4\ge 0, \dots, vbu_1-u_4\ge 0,\dots\}$.
We then filter $\mathit{Cons}$ to $\mathit{Cons}^{*}$ where
\[
\mathit{Cons}^{*} = \{c\in \mathit{Cons} \mid \phi'' \land x-u_4=0 \land y-u_5=0 \land a2-u_6\land \dots \land u_3-u_1=0\models_{\textit{LRA}} c\}.
\]
What $\mathit{Cons}^{*}$ gives is a set of equalities and inequalities that we will add to the cone of polynomials. However, we have to take one more step before we add the inequalities. For this example, $\mathit{Cons}^{*} = \{e\ge0, \dots, vbu_1-u_4, \dots\}$.

\subsection{Taking Products}\label{Se:takingProducts}
Before we add inequalities to the cone, we ``take products,'' which is a strengthening step indicated as step (5) in \figref{coneSat}.
The process of taking products is one of the main reasons our method gives interesting answers, and it is what leads to the main expense of the overall method.
Suppose that from step (4) we obtain the inequalities $x\ge 0, y\ge0, z\ge 0$.
In non-linear arithmetic, from these inequalities we have $x^2\ge0, xy\ge0, xz\ge0,y^2\ge 0, yz\ge0, z^2\ge 0, x^3\ge 0$, etc.
Moreover, all of these ``product'' inequalities are not members of $[x,y,z]$. We could strengthen the cone by adding all of these product inequalities; however, the set of all of these products is infinite.

In our implementation, we heuristically ``cut-off'' the depth of products at some parameterized value $N$.
That is, we assign each inequality $w\ge 0$ in the cone a depth $i$, which we denote by $w\ge_i 0$.
Newly discovered inequalities, i.e., the ones produced from step (4), have a depth of 1.
Product inequalities can be generated by the rule $w\ge_i 0$ and $z\ge_j 0$ yields $wz\ge_{i+j} 0$.
When we add inequalities to the cone, we make sure to add all products that have a depth less than or equal to $N$.
For example, suppose that the polyhedral cone $[x, y, x^2, xy, y^2]$ corresponds to the following inequalities with indicated depths
$\{x\ge_1 0, y\ge_1 0, x^2\ge_2 0, xy\ge_2 0, y^2\ge_2, 1\}$, $N=2$, and we have the newly discovered inequality $z\ge_1 0$.
We make sure to take all products within the new inequalities (\{$z^2\ge_2 0$\}), as well products with the polyhedral basis ($\{xz \ge_2 0, yz\ge_2 0\}$), to obtain the new inequalities $\{z\ge_1, z^2\ge_2 0, xz\ge_2 0, yz\ge_2 0\}$.
Thus, after taking products and adding the results to the cone of polynomials we would have $C= \langle p_1,\dots,p_k\rangle + [x, y, x^2, xy, y^2, z, z^2, xz, yz, 1]$.

For the running example we use a saturation depth of $N=3$, and would generate many inequalities from $\mathit{Cons}^{*}$ from~\sectref{consFinding}. Generated inequalities would include $e\ge_1 0, e^2\ge_2 0, e^3\ge_3 0, b\ge_1 0, eb\ge_2 0, vbu_1\ge_3 0$, and many more, all of which would be added to the cone of polynomials.

\subsection{Putting it All Together}
For the running example, after going through one round of saturation, we have the following:\footnote{
To simplify the presentation,
we have removed some clauses from $\phi''$ that are no longer useful.}
\begin{align*}
    \phi''\defeq\ &(vbu_1\ge 0\implies u_4\ge 0)\land \dots\\
    C \defeq\ &\langle x-u_4, y-u_5, a2-u_6, e'-e-a, b'-b-a2, eu_1-1, eu_2+au_2-1, u_3-u_1\rangle + \\
              &[e, e^2, e^3, b, \dots, vbu_1, \dots, 1]\\
    TM \defeq\ &\{u_1\mapsto e^{-1}, u_2 \mapsto e'^{-1}, u_3\mapsto e^{-1}, u_4 \mapsto \floor{vbu_1}, u_5 \mapsto \floor{vb'u_2}, u_6\mapsto \floor{abu_3}\}
\end{align*}
However, going through the saturation steps again will generate even more information. For example, in $\phi''$ we have $vbu_1 \ge 0 \implies u_4 \ge 0$.
When we performed consequence finding in step (4), $u_4$ was not a true consequence, because it was not \emph{linearly} implied by $C$ or $\phi''$.
However, by taking products of the inequalities $v\ge0$, $b\ge0$, and $u_1\ge 0$, we now have $vbu_1$ as a basis polynomial in the polyhedral cone. Therefore, now $u_4\ge 0$ \emph{is} linearly implied by $C$ and $\phi''$, so running through the steps again would establish $u_4\ge 0$.
Similarly, it may be possible for new equations to be generated in steps (2) or (4), as well.

The saturation process starts with an ``empty'' cone
$C_0 = \langle 0 \rangle + [1]$.
Then, at each step of saturation a stronger cone is produced that is still implied by the original formula.
That is, starting from $C_0$ a sequence of cones is produced $C_0, C_1, \dots, C_n$ where $C_i$ is created by running one step of saturation on $C_{i-1}$.
Each cone is implied by the original formula together with the foreign-function map. Furthermore, $C_{i-1} \subsetneq C_i$ for $i\le n$.
\begin{theoremrep}
For each $i$. We have for every $c\in C_i$,
\[
\phi \land \bigwedge_{u\mapsto f(p)\in TM} (u = f(p)) \models_{UFOF} c\ge 0.
\]
Where $\models_{UFOF}$ denotes entailment w.r.t. the theory of ordered fields with uninterpreted functions.
\end{theoremrep}
\begin{appendixproof}
Here we give a proof sketch of the idea. First note that by \lemref{coneSound}, to show every element of the cone is sound, we only need to show that each element of the basis of the ideal is an implied equation and each element of the basis of the polyhedral cone is an implied inequality. We prove the theorem by induction on $i$. For $C_0 = \langle 0\rangle + [1]$, we have the only elements of $C_0$ are positive scalars.

For the inductive step, suppose the theorem holds for $C_i$. Saturation creates $C_{i+1}$ from $C_i$ in a few different ways. To show the theorem we must consider each possibility. Step (2) (closure) can add new equations to the ideal. In this step we add the equation $w-w'$ to the ideal when we have $w\mapsto f(p)$ and $w'\mapsto f(p')$ and $p=p'$ is implied by the cone $C_i$. By the inductive hypothesis $p=p'$ is implied by the formula and function map, so by the function axiom $p=p'$ entails $f(p) = f(p')$ we have $w=w'$ entailed by the formula and function map. Another option to create $C_{i+1}$ is through consequence finding. We query an SMT solver with a formula equivalent to $\phi \land \bigwedge_{u\mapsto f(p)\in TM} (u = f(p)) \land C_i\models_{LRA} cons$ which implies $\phi \land \bigwedge_{u\mapsto f(p)\in TM} (u = f(p)) \land C_i\models_{UFOF} cons$. For any consequence $cons$ that is an inequality we also ``take products'', which is also sound with respect to ordered fields. These are the ways new equalities and inequalities get added to $C_i$, and so the resulting cone $C_{i+1}$ is also sound. Thus by induction the theorem holds.
\end{appendixproof}

What is key, though, is that saturation will eventually terminate. That is, $C_n = C_{n+1}$ for some $n\in \mathbb{N}$. The reason is two-fold.
With respect to inequalities,
because we limit inequalities with a saturation bound, there are a finite set of potential inequalities (coming from the finite formula) and inequality products because products from a finite set give a finite set.
With respect to equalities,
we can appeal to Hilbert's basis theorem (see~\sectref{background}) and say that
the set of equalities must eventually stabilize
because every polynomial ideal is finitely generated.

For the running example, running the saturation procedure until it stabilizes---using a saturation bound of 3---produces a cone $C = \langle x-u_4, y-u_5, \dots\rangle + [vbu_1-u_4,vu_2u_6+vbu_2 - vbu_1 + vu_2, u_5, vu_2u_6-vbu_2 + 1, \dots, 1]$.
Reducing $x-y$ using the procedure from \sectref{coneRed} gives the upper bound $vu_2+1$.

%% file: figures/consat.tex
\begin{tikzpicture}[
  node distance = 0.7cm,
  mynodes/.style = {very thick, draw=black!50,top color=white,bottom color=black!20, font = \ttfamily, font=\small, minimum width = 2.5cm},
ideal/.style = {mynodes, diamond, aspect=2, rounded corners},
  tmap/.style = {mynodes,rectangle, rounded corners=3mm},
  pset/.style = {mynodes, rectangle},
  eqR/.style = {rounded corners, draw, top color=blue!15, bottom color=blue!25},
  cone/.style = {rounded corners, draw, dashed, top color = red!15, bottom color = red!25},
  textnode/.style={font=\small}
]
\begin{pgfonlayer}{foreground layer}
  \node (tmap) [tmap] {term map};
  \node (ideal) [ideal,below=of tmap] {$\langle$ eq's $\rangle$};
\end{pgfonlayer}

\node (eqR) [eqR, fit=(ideal) (tmap),opacity=0.8] {};
\node (eqRLabel) [above,textnode] at (eqR.north) {};
%{Eq Rewriter};
   
\node (ineqs) [tmap, below =of ideal] {inequalities};
%\begin{pgfonlayer}{background layer}
  \node (cone) [cone, fit=(ideal) (ineqs),label=above:Cone,opacity=0.8,minimum width=3.2cm, textnode] {};
%\end{pgfonlayer}
\node (ineqs1) [tmap, below =of ideal] {inequalities};

\node(newIneqs) [left= 0.75cm of ineqs1,textnode] {new ineq's};
\draw [->] (newIneqs.east) -- (ineqs.west);

\node (newEqs) [above=1mm of newIneqs,textnode] {new eq's};
\draw [->] (newEqs.north) -- (newEqs.north |- ideal.west) -- node (addL) [above,textnode] {\hspace{-6ex} (1) Add eq's} (ideal.west);
%\draw [->] (newEqs.east) -- node (addL) [above] {\hspace{-4ex} \small(1) Add eq's} (ideal.west);

\draw [<->] (ineqs.north) -- node [right,textnode] {(3) Reduce} (eqR.south);

\draw [->] (eqR.30) -- 
           ++(right:0.25cm) -- node[right,textnode] {(2) Closure} 
           ++(up:0.75cm) -- 
           ++(left:0.5cm) -- 
           (\currentcoordinate |- eqR.north);

\node (extract) [right=0.8cm of ineqs.east, text width = 2.4cm, align=center,textnode] {(4) Extract eq's and consequences};
\node (impls) [above= 1cm of extract, text width = 2.5cm, align=center,textnode] {Potential consequences};

\draw [->] (impls.south) -- (extract.north);
\draw [->] (extract.250) -- ++(down:0.7cm) -- node[above,textnode] {(5) Take products} (\currentcoordinate -| newIneqs.south) -- (newIneqs.south);

\path (newEqs.west) -- ++(left:0.3) coordinate (c);
\draw[->] (extract.290) -- ++(down:0.9cm) -- (\currentcoordinate -| c) -- (c) -- (newEqs.west);
%\draw [<-] (newEqs.west) -- ++(left:0.5) -- ++(down:2.5cm) -- (\currentcoordinate -| extract.290) -- (extract.290);

\draw [->] (cone.331) -- (\currentcoordinate -| extract.west);

\end{tikzpicture}

%% file: effective_degree.tex
\section{Effective degree order}\label{Se:effectiveDeg}
The methods of \sectref{reduction} can reduce a polynomial $t$ by a cone of polynomials $C$ with respect to an arbitrary monomial order. Thus, a user can use the results of saturation, a purified term to rewrite $t$, a cone of polynomials $C$, and a foreign-function map $TM$, to create an arbitrary monomial order for some downstream task. However, determining an appropriate order is a challenging task. In this section we present the \emph{effective-degree order} which is the monomial order we use in \Tool.

Our definition for effective degree includes a set of variables $W$, specified by the user, which indicate variables to keep. That is, by specifying a set $W$ the user is indicating that any term containing a variable not in $W$ is worse than a term containing only $W$ variables. In this way, the set $W$ encodes a ``variable restriction'' on the preference of terms. Incorporating variable restriction into a traditional monomial order is straightforward. However, in our setting we have the additional challenge of function symbols in the foreign-function map $TM$. Suppose we have the assignment $x\mapsto f(p)$ in $TM$. If $p$ contains only $W$ variables, then in the monomial order the variable $x$ should also be thought of as referring to only $W$ variables. However, it may be the case that $p$ contains some variables \emph{not} in $W$, but there could be another polynomial $p'$ with $p=p'$ and $p'$ \emph{does} contain only $W$ variables. For example, consider the assignment $u_2\mapsto e'^{-1}$ in the running example, and let $W=\{a, b, e, v\}$. $e'$ is \emph{not} in $W$, but we have $e' = e+a$ implied by the ideal. In other words, we have $u_2\mapsto (e+a)^{-1}$, and now $u_2$ refers to a function containing only $W$ variables. 

To solve this challenge we must first ``reduce'' the foreign-function map $TM$ using the ideal in $C$. The idea is that we have an initial definition for effective degree with variable restriction. We then reduce each polynomial $p$ in each assignment in $TM$. Each reduction may rewrite $p$ towards another $p'$ which is lower in effective degree and thus may only contain $W$ variables. We then update the effective degree order and repeat until the process stabilizes. \footnote{In our experiments we always use an effective-degree order so this reduction step occurs during closure, discussed in \sectref{closure}, and not as a separate step.}

Let $W$ be a set of variables, $TM : Y\rightarrow \text{Term}$ a foreign-function map. Let $m$ be a monomial over variables $X$ with $Y\subseteq X$. The \emph{effective-degree} of a variable $x$ denoted $\effdegmw{TM}{W}{m}$ is a pair of natural numbers and is defined recursively as follows:
\[
\begin{gathered}
\effdegmw{TM}{W}{x} \defeq \begin{cases}
                                 (0, 1) & \text{if } x\notin TM \text{ and } x\in W\\
                                 (1, 0) & \text{if } x\notin TM \text{ and } x\notin W\\
                                 \LM(p) & x\mapsto f(p) \in TM
                               \end{cases}\\
    \begin{array}{c@{\hspace{4ex}}c}
        \effdegmw{TM}{W}{n} \defeq (0, 0) & \effdegmw{TM}{W}{st} \defeq \effdegmw{TM}{W}{s} + \effdegmw{TM}{W}{t}
    \end{array}
  \end{gathered}
\]
In the above $+$ is taken pointwise, and $\LM(p)$ is $\max(\effdegmw{TM}{W}{m_1},\dots,\effdegmw{TM}{W}{m_n})$ with $\max$ taken in lexicographic order w.r.t.\ the monomials $m_1,\dots,m_n$ of $p$. To make $\effdegmw{TM}{W}{-}$ a total order we break ties between $\effdegmw{TM}{W}{x}$ and $\LM(p)$ for $x\mapsto f(p) \in TM$ by taking $x\MOGE \LM(p)$.

\begin{example}
Consider the un-reduced map $TM$ from earlier, and let $W=\{a, b, e, v\}$
\[
TM \defeq\ \{u_1\mapsto e^{-1}, u_2 \mapsto e'^{-1}, u_3\mapsto e^{-1}, u_4 \mapsto \floor{vbu_1}, u_5 \mapsto \floor{vb'u_2}, u_6\mapsto \floor{abu_3}\}
\]
$\effdegmw{TM}{W}{u_5} = \effdegmw{TM}{W}{v} + \effdegmw{TM}{W}{b'} + \effdegmw{TM}{W}{u_2} = (0, 1) + (1,0) +(1, 0) = (2, 1)$. Now consider the reduced map $TM'$, using the ideal of $C$:
\[
TM' \defeq\ \{u_1\mapsto e^{-1}, u_2 \mapsto (e+a)^{-1}, u_3\mapsto e^{-1}, u_4 \mapsto \floor{vbu_1}, u_5 \mapsto \floor{v(b+u_6)u_2}, u_6\mapsto \floor{abu_3}\}
\]
$\effdegmw{TM'}{W}{u_5} = (0, 5)$.
\end{example}
% \begin{lemmarep}
% For a well-defined map $TM$ $\effdegmw{TM}{W}{-}$ is a monomial order.
% \end{lemmarep}
% \begin{appendixproof}
% \john{TODO}
% \end{appendixproof}
\begin{lemmarep}
Reducing a foreign-function map $TM$ using effective degree eventually stabilizes.
\end{lemmarep}
\begin{appendixproof}
We must check two things to show the lemma. First, we need to make sure that effective degree remains well-defined during the reduction process. Namely, if the map $TM$ contains a cycle then effective degree is \emph{not} well-defined. Let $>$ be a partial order on variables of $TM$ such that $u > u'$ if $u'\in \textit{vars}(p)$ with $u\mapsto f(p)\in TM$. If such a partial order exists the map is well-defined. In the effective degree order we require $u\MOGE \LM(p)$. Thus, if $u'\in p$ we must have $u\MOGE u'$. Moreover at every step of reducing the map we have $u\MOGE \LM(p)\MOGE\LM(\textbf{red}_B(p))\MOGE u''$ for every $u''$ in $\textbf{red}_B(p)$. Thus, assuming $TM$ is well-formed to begin with, at every step of reduction we can take the partial order that shows $TM$ is well-defined to be the effective degree order.

To show termination, we define the effective degree of $TM$ as $\max_{u\mapsto f(p)\in TM} \effdegmw{TM}{W}{u}$. Consider reducing the map with respect to some basis $B$ to create some map $TM'$. By \lemref{reduceReduces}, we have that $\effdegmw{TM}{W}{u} = \LM(p)\MOGE\LM(\textbf{red}_B(p)) = \effdegmw{TM'}{W}{u}$ for each $u\mapsto f(p)$ in $TM$. Thus the effective degree of $TM'$ is smaller than the effective degree of $TM$. Moreover, $\effdegmw{TM}{W}{x}\in \mathbb{N}\times \mathbb{N}$ is well-ordered with minimal element $(0, 0)$. Thus, termination is guaranteed.
\end{appendixproof}
In~\sectref{experiments} we use effective degree to mean the effective degree w.r.t. a reduced map $TM$.

%% file: experiments.tex
\section{Experiments}\label{Se:experiments}
We implemented our technique in a tool called $\Tool$,\footnote{We will make the code and benchmarks publicly available.} 
handling arithmetic terms with floors and divisions.\footnote{
We used in saturation (\sectref{saturation}) the equality axiom for $e \cdot \frac{1}{e}$ mentioned in~\sectref{overview}, and the inequality axioms $\forall e. \, e - 1 \leq \floor{e} \leq e$, $\forall e. \, e \geq 0 \implies \frac{1}{e} \geq 0$, and $\forall e. \, e \geq 0 \implies \floor{e} \geq 0$.
}
We use Z3 \cite{z3} for SMT solving, and the FGb library \cite{fgb} \footnote{\url{https://www-polsys.lip6.fr/~jcf/FGb/index.html}.} for Gr\"obner-basis calculations.
Our evaluation of $\Tool$ targets the following questions, grouped in two categories:
\begin{enumerate}
    \item \textbf{Performance}: These questions address the run-time behavior of $\Tool$, and its scalability, and how the time breaks down by component.
    We study:
    \begin{enumerate}
        \item How much \textbf{overall time} does it take for $\Tool$ to produce a bound?
        \item How does the overall time \textbf{break down} into the time spent in the \textbf{saturation step} and in the \textbf{reduction step}? %\yf{other steps not represented here, or should always sum to the total?}
        %\twr{To focus just on those two components, we could say, ``What fraction of the time is spent in the saturation step? What fraction in the reduction step?}
        \item What is the \textbf{size of the (representation of the) cone} that $\Tool$ generates---how many \textbf{equalities} are produced to form the ideal, and how many \textbf{inequalities} over how many distinct \textbf{monomials} are produced to form the polyhedral cone? (The number of monomials produced gives the number of dimesions of the corresponding polyhedron and the number of inequalities gives the number of constraints of the polyhedron.)
        \item How does~\algref{polyReduce} compare with the naive method of polyhedral-cone reduction that is based on linear programming, outlined in~\sectref{lpRedux}.\footnote{In our experiments we used Z3's Optimize module to solve the multi-objective linear program.}
        \item How does $\Tool$ \textbf{scale} as the \textbf{saturation bound} is increased?
    \end{enumerate}
    
    \item \textbf{Bound quality}: These questions examine the output bound that $\Tool$ synthesizes.
    %We study:
    \begin{enumerate}
        \item Is the bound \textbf{optimal} w.r.t.\ the effective-degree order?
        \item Is the bound \textbf{desirable}---tight yet meaningful?
        \item What \textbf{saturation depth} suffices for $\Tool$ to synthesize desirable bounds?
        %generates plateau? \yf{might need to increase the saturation bound beyond what we currently have in the graph to know that it actually plateaus; we do know that it hits what we consider an optimal bound}
        % \zak{Converge means the bound does not change as we increase depth?  I don't think we can test for this.  Say "until the bounds plateau?"}
    \end{enumerate}
\end{enumerate}

% Based on the same benchmarks, we also discuss our experience with \textbf{bound desirability}: Is the bound we obtain considered tight yet meaningful for the problem at hand?
% \yf{Not sure how to evaluate bound optimality, maybe discuss it together with what we can say about desirability.}

We investigated these questions using a set of benchmarks that we collected from Solidity code provided to us by industry experts in smart-contract verification, which we modelled in our tool as a set of equational and inequational assumptions over initial and intermediate variables in the program.
Overall, we find that $\Tool$ is able to produce optimal, meaningful bounds in a few seconds in nearly all the benchmarks.
We begin with a brief description of the benchmarks.

\subsection{Benchmarks}
We briefly describe each of the problems that we considered in our evaluation.

\paragraph{Elastic}
This is the example that was described in detail in~\sectref{overview}.

% \paragraph{Fixed-point arithmetic: multiply and divide}
\paragraph{Fixed-point arithmetic}
Here, non-integer numbers are represented by multiplying them with a scaling factor $\textit{sf}$, i.e., a rational number $x \in \mathbb{Q}$ is approximately represented by rounding $x \cdot \textit{sf}$ to an integer.  Multiplication and division need to correct for the double occurrence of the scaling factor, dividing by $\textit{sf}$ upon multiplication and multiplying by $\textit{sf}$ upon division. All these divisions are \emph{integer} divisions and so not exact.
We are interested in what happens when a number represented by $a \in \mathbb{Z}$ is multiplied and then divided by the same number represented by $b \in \mathbb{N}$, that is, in the term
$
    t = \floor{\frac{\floor{\frac{ab}{\textit{sf}}} \cdot \textit{sf}}{b}},
$
and seek a bound over any of the variables present, under the inequational assumptions $b \geq 0, \textit{sf} \geq 0$.
Note the nested structure of floor terms.
%\zak{Typesetting $\frac{a}{b}$ and $a/b$ right next to eachother looks a little odd if they're intended tp mean the same thing}

% \paragraph{Fixed-point arithmetic: divide and multiply}
% This is similar to the previous benchmark, except that the operation first divides and then multiplies:
% $
%     t = \floor{\frac{\floor{\frac{a \cdot \textit{sf}}{b}} \cdot b}{\textit{sf}}}.
% $
% \yf{actually this is identical to the previous example under switching $b,\textit{sf}$. should we keep just one?}

% The upper bound is optimal (because a constant bound is impossible) and tight (if all divisions are exact, $t_1 = t_2 = a$).
% The lower bounds are good when $b,\textit{sf}$ are approximately of the same order of magnitude.
% We asked the industry verification expert about the the lower bound we produce, and in response (after correcting some error in their derivation) they produced the bound $a - \frac{ab \bmod \textit{sf}}{b} \leq t_1$. The closest bound in the language for terms that we use is $a - \frac{\textit{sf-1}}{b} \leq t_1$ (using the inequality $ab \bmod \textit{sf} \leq \textit{sf}-1$), which is very close to our bound.

\paragraph{Manual price}
\newcommand{\texttstroke}{\tilde{t}}
There is an auction with a price that decreases linearly with time. The price in time $\texttstroke \in [\textit{beginTime},\textit{endTime}]$ is computed by
\begin{equation*}
    \textit{drop}(\texttstroke) = \floor{\frac{\textit{startPrice} - \textit{minimumPrice}}{\textit{endTime} - \textit{startTime}}},
    % \\
    \quad
    \textit{price}(\texttstroke) = \textit{startPrice} - \textit{drop}(\texttstroke) \cdot (\texttstroke - \textit{startTime}).
\end{equation*}
To show that the price always resides in the expected interval, we want bounds on the term
$
    t = \textit{price}(\texttstroke),
$
under the inequational assumptions $\textit{startTime} \leq \texttstroke \leq \textit{endTime}$, $\textit{minimumPrice} \leq \textit{startPrice}$, and over any of the variables present. 

\paragraph{Manual-price monotonicity}
In the context of the previous benchmark,
to show monotonicity of the price with time, we are interested in an upper bound on the term
$
    t_2 = \textit{price}(\texttstroke_2) -\textit{price}(\texttstroke_1)
$
under the inequational assumptions $\textit{startTime} \leq \texttstroke_1 \leq \texttstroke_2 \leq \textit{endTime}$, $\textit{minimumPrice} \leq \textit{startPrice}$, and over any of the variables present.

\paragraph{Token}
In one implementation of an ``elastic'' smart contract by TrustToken\footnote{\url{https://github.com/trusttoken}}, one property of interest is that a client cannot increase their gain by splitting transactions; specifically, that $\textit{balanceSplit}$, the balance of a client after \cinline{withdraw(x);withdraw(x)}, is not greater than $\textit{balanceJoined}$, the balance after executing \cinline{withdraw(2x)} from the same initial state. Each \cinline{withdraw} operation modifies the client's balance, including a fee for the operation, as well as the total supply held by the contract, in ways that involve several multiplications and divisions with another quantity called \cinline{funds}. We are interested in bounds for the term
$
    t = \textit{balanceSplit} - \textit{balanceJoined},
$
under inequational assumptions that the client's balance and total supply are large enough to withdraw $2x$, and that \cinline{funds} is non-negative, over the variables $x$, the initial values of the client's balance and supply, and \cinline{funds} (but not over value of intermediate variables).

% The upper bound is optimal and almost tight---the true bound seems to be $1$.
% \yf{note this:}
% The lower bound is novel and interesting, but whether it is optimal is not known.

\paragraph{NIRN}
In a different implementation of an ``elastic'' smart contract by NIRN\footnote{\url{https://github.com/indexed-finance/nirn}}, the property of interest is that when a client deposits an amount and receives a corresponding amount of shares (see~\sectref{overview}), the number of shares does not vary much at different times for the same amount. Specifically, the question is what happens under an execution of \cinline{shares1 = deposit(x); shares2 = deposit(x)} to the term
$
    t = \textit{shares1} - \textit{shares2},
$
under appropriate inequational assumptions that ensure that deposit operations execute successfully.
%a successful operation \twr{Break the sentence here?  The next part is about what we want $\Tool$ to do.} and
We would like to find a bound for $t$ defined over the variable $x$ and the initial values alone (not over values of intermediate variables).
Each \cinline{deposit} operation modifies the total supply and the client's balance, in a calculation that is more involved than in the benchmark \textit{Token}, described above.

% which is optimal and to the best of our knowledge also tight.
% \yf{note} However, we are not able to generate a meaningful lower bound for this term, producing a complex term with several floor functions even with a saturation bound of $3$.

\subsection{Performance}
\input{basic_table.tex}
\Cref{Ta:Rewriting} displays the running time of \Tool on each of the benchmarks with saturation depth $3$.
Time measurements are averaged over 5 runs, running on an Ubuntu 18.04 VM over an X1 Carbon 5th Generation with an Intel(R) Core(TM) i7-7500U CPU @ 2.70GHz 2.90 GHz processor.

\subsubsection{Overall time}
In the majority of our benchmarks, $\Tool$ is able to produce the bound in less than a few seconds. The only exception is the NIRN benchmark, where the upper bound requires about a minute and a half, which we attribute to the larger number of floor and division terms and accordingly a larger resulting cone. %\zak{is the effective degree of the terms in the input a critical factor?}
%, and the lower bound does not terminate in less than \yf{several hours}. \yf{need to decide whether we run this benchmark with more inequational assumptions, where the tool doesn't terminate, or with less, where the tool terminates with a useless bound}
% \zak{Where does the time go?}\yf{not sure I understood} \zak{Also, it's a bit surprising to see the difference between nirn and tokent, given that these problems are roughly the same size.  Why does nirn blow up to 10x as many monomials and inequations?  Is the (effective) degree of the terms in the input a critical factor?}\yf{could this be the number of floor terms in the equational assumptions? actually most of the token equational assumptions correspond to simple substitution}
% We attirubte the greater challenge $\Tool$ has with NIRN to the larger number of floor terms in the assumptions, which (through the instatination of floor axioms and subsequent saturation with products) a much larger cone.

\subsubsection{Breakdown}
In most benchmarks, saturation-time is slightly larger than reduce-time. The only exception is NIRN, where reduce-time dominates. This may indicate that the time to reduce increases more steeply then the time saturate as the cone size increases.
%as the cone gets larger, the time to create the cone increases, but the time to use the cone increases more steeply. 
This is also apparent in the trends when saturation depth is increased (shown in~\refappendix{SaturationDepthTable}).

\subsubsection{Cone size}
The size of the cone affects the running time of both saturation and reduction.
The number of equalities in the ideal grows with the number of equational assumptions in the input, although not steeply.
The number of inequalities in the cone grows rather steeply with the number of inequational assumptions, and especially with the number of integer divisions---these growth patterns are likely because floor and division terms trigger the instantiation of inequational axioms,
which in the saturation process further amplifies the number of inequalities because of taking products
of inequational assumptions and inequational terms from axiom instantiations.

\subsubsection{Reduction with Linear Programming}
In all benchmarks, performing the reduction step using linear programming is significantly slower than using our local projection method, often in one or more orders-of-magnitude, justifying our use of local projection for reduction.

\subsubsection{Scalability with Saturation Depth}
\begin{wrapfigure}{R}{0.5\textwidth}
\begin{minipage}{0.48\textwidth}
\vspace{-0.8cm}
\begin{figure}[H]
    \centering
    \includegraphics[scale=0.45]{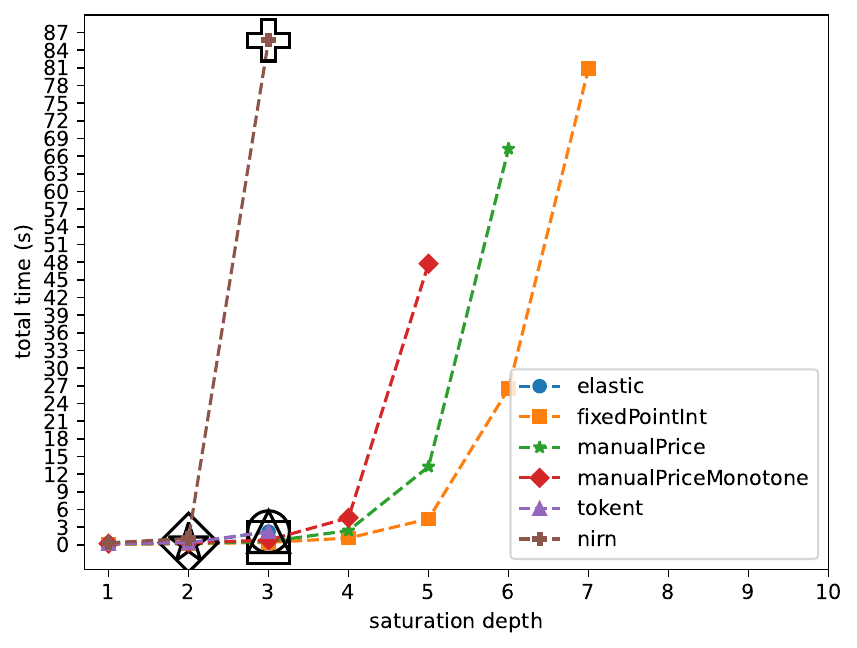}
\vspace{-0.5cm}
    \caption{\small Total running time as the saturation depth is increased. Missing points indicate a time greater than the timeout of 90 seconds.
% \twr{90 seconds seems rather short.
% I realize that we are dealing with an exponentially growing function, but I would have thought 10-15 minutes would be more reasonable.
% }
    For each benchmark, the depth that suffices for both the upper and lower bounds to be the ones given in
    \tableref{EvalQuality} is marked by a larger, black marker of the same type as the line plot for the benchmark. % for example, in token, 2 is enough for the upper bound but 3 is necessary for a nice lower bound
    }
    \label{Fi:SatScalability}
\vspace{-0.5cm}
\end{figure}
\end{minipage}
\end{wrapfigure}
% \begin{figure}[t]
%     \centering
%     \includegraphics[scale=0.65]{saturation_scalability.pdf}
%     \caption{\small Total running time as the saturation depth is increased. Missing points indicate a time greater than the timeout of 90 seconds.
% % \twr{90 seconds seems rather short.
% % I realize that we are dealing with an exponentially growing function, but I would have thought 10-15 minutes would be more reasonable.
% % }
%     For each benchmark, the depth that suffices for both the upper and lower bounds to be the ones given in
%     \tableref{EvalQuality} is marked by a larger, black marker of the same type as the line plot for the benchmark. % for example, in token, 2 is enough for the upper bound but 3 is necessary for a nice lower bound
%     }
%     \label{Fi:SatScalability}
% \end{figure}
\figref{SatScalability} shows how the overall time of $\Tool$ changes as the saturation depth is increased, using a timeout of 90 seconds (each data point averaged over 3 runs).
We see that running time
increases steeply, which is expected because the number of product terms grows exponentially with the saturation depth. 
%However, the default saturation bound of $3$ performs well (and suffices---see~\sectref{EvalQualitySatDepth}) in all examples \yfchanged{but the NIRN lower bound}. 
We are pleasantly surprised by the fact that $\Tool$ tolerates even larger saturation depths for some of the benchmarks.
However, saturation depth of 8 causes $\Tool$ not to terminate within the time limit in any of the benchmarks.
Further data on the cone size and the run-time breakdown as the saturation depth is increased until a 10 minutes timeout appear in~\refappendix{SaturationDepthTable}.
\begin{toappendix}
\section{Detailed Statistics from Saturation Depth Experiment}
\label{Se:SaturationDepthTable}
\input{sat_size_table}
The cone size and run-time breakdown of $\Tool$ as the saturation bound is increased in each benchmark, until a timeout of 10 minutes.
\end{toappendix}
% \twr{Presumably the cone gets bigger and bigger.
% I think it would be interesting to see a version of \tableref{Rewriting} in which we give the cone-size and running-time statistics as the saturation bound increases.
% }
% \twr{Do we ever get a different final bound as we increase saturation depth?}
% \yf{beyond the bounds in~\tableref{EvalQuality}, I think no. I'll check the logs. Though I don't have data for token and nirn beyond saturation depth 3.}

\subsection{Bound Quality}
\begin{table}[t!]
	\centering
	\small
	{\small \caption{\label{Ta:EvalQuality}
            {\small Comparison of the bounds that $\Tool$ generates for input term $t$ (saturation depth=3) and the bounds curated by a human expert. $\star$ indicates a result that is too large to include in the table.
            ed shows the (second component of the) effective-degree of the bound (the first component is 0).
	}}}
\begin{tabular}{|| l | c | c | c | c | c ||}
\hline
\multirow{2}{*}{Benchmark name} & \multirow{2}{*}{$t$} & \multicolumn{2}{c |}{$\Tool$} & \multicolumn{2}{c ||}{Expert} \\
& & bound & ed & bound & ed \\
\hline\hline
\multirow{2}{*}{elastic} & \multirow{2}{*}{$x-y$} & $\leq {\frac{v}{e+a}} + 1$ & 2 & $\leq \floor{\frac{v}{e+a}} + 1$ & 2 \\
& & $\geq -1$ & 0 & $\geq 0$ & 0 \\
\hline

\multirow{2}{*}{fixedPointInt} & \multirow{2}{*}{$\floor{\frac{\floor{\frac{ab}{\textit{sf}}} \cdot \textit{sf}}{b}}$} & $\leq a$ & 1 & $\leq a$ & 1 \\
& & $\geq a - 1 - \frac{\textit{sf}}{b}$ & 2 & $\geq a - \frac{\textit{sf-1}}{b}$ & 2 \\
%& \multirow{2}{*}{$\floor{\floor{\frac{a \cdot \textit{sf}}{b}} \cdot b \ / \ {\textit{sf}}}$} & $\leq a$ & $ \leq a$ \\
%& & $\geq a - 1 - \frac{b}{\textit{sf}}$ & n.a. \\
\hline

\multirow{2}{*}{manualPrice} & \multirow{2}{*}{$\textit{price}(\texttstroke)$} & $\leq \textit{startPrice}$ & 1 & $\leq \textit{startPrice}$ & 1\\
& & $\geq \textit{minimalPrice}$ & 1 & $\geq \textit{minimalPrice}$ & 1 \\
\hline

manualPriceMonotone & $\textit{price}(\texttstroke_2) - \textit{price}(\texttstroke_1)$ & $\leq 0$ & 0 & $\leq 0$ & 0 \\
\hline

\multirow{2}{*}{token} & \multirow{2}{*}{$\textit{balanceJoined} - \textit{balanceSplit}$} & $\leq 2$ & 0 & $\leq 1$ & 0 \\
& & $\geq -1/2 - 1/2 \cdot \textit{funds}$ & 1 & n.a. & -- \\
\hline

\multirow{2}{*}{nirn} & \multirow{2}{*}{$\textit{shares1} - \textit{shares2}$} & $\leq 1.88$ & 0 & n.a. & -- \\
& & $\star$ & 5 & n.a. & -- \\
\hline
\end{tabular}
\end{table}

\tableref{EvalQuality} presents the bounds we obtain for the benchmarks with saturation depth $3$, as well as the bounds generated manually by a verification expert (where available). 
$\star$ indicates that $\Tool$ has generated a bound we deem as uninformative.
As explained in~\sectref{reduction}, there are potentially multiple optimal upper-bounds. Moreover,
the technique can discover several upper bounds of the same effective-degree in one run of \algref{conjBound}; in $\Tool$ we simply return multiple bounds in that case. In the table, this arose in fixedPointInt (5 upper bounds), where it was easily to select the most appealing of them (the other involving additional added terms); and NIRN (2 lower bounds), where both bounds were uninformative, as indicated in the table by $\star$. The latter case we discuss separately, below in~\sectref{Desirability}.

\subsubsection{Optimality}
The bounds that $\Tool$ produces are nearly optimal w.r.t.\ the effective-degree order.
The lower bound for \emph{elastic}, and the upper bound for \emph{manual-price monotonicity}, \emph{token}, and \emph{NIRN} are all constant, which is always optimal.
The upper bounds for \emph{fixed point} and \emph{manual price} are also optimal, consisting merely of one variable.
(These two results are optimal because there is no valid constant bound;
moreover, there is no other bound consisting of a single variable that is valid.)
%, seeing that no non-constant bound is valid \twr{What does the previous phrase mean?}, and no other variable is a valid upper bound \yfchanged{(internally, $\Tool$ uses an effective-degree order with an arbitrary order on the variables)}.
It is harder to evaluate whether the remaining bounds are close to optimal w.r.t.\ the effective-degree order; however, they have the same effective-degree as the expert bounds, where the latter are available.
%\yfchanged{(Increasing the saturation depth the way we report in~\refappendix{SaturationDepthTable} does not alter the resulting bounds.)}

%\yf{is there a set of bounds that would convince us of our bound's optimality if we checked them in Z3?}
%\zak{I don't think that we can verify optimality using Z3.  My recollection is that Z3 lacks support for quantifiers for \textit{linear} integer+real arithmetic, let alone non-linear.}

\subsubsection{Desirability}
\label{Se:Desirability}
As~\tableref{EvalQuality} demonstrates, the bounds that $\Tool$ computes match or nearly match the bounds produced by a domain expert, and differ in constant or nearly-constant (e.g.\ $\frac{1}{b}$) terms. We attribute these differences to the challenge of inequality reasoning inside a function, which we discussed in~\sectref{UnderFloorRewrite}.
%of reducing under the floor function, which we discuss in~\Cref{Rem:UnderFloorRewrite}.}\twr{Wait -- this is \Cref{Rem:UnderFloorRewrite}!}

% \begin{changebar}
% \begin{remark}
% \label{Rem:UnderFloorRewrite}
% \yf{suggesting to put this in the technical part about saturation/axioms}
% One interesting limitation of our approach is that we instantiate inequational axioms only over terms that exist in the input formula. For example, if $\tilde{t}$ is not present in the input, then we do not include in the cone that fact that $\floor{t} \leq \floor{\tilde{t}}$ even if we generally do instantiate the monotonicity axiom for $\floor{\cdot}$ \yfchanged{and can derive the fact $t \leq \tilde{t}$}.
% % \twr{Confusing because the paragraph never indicates that we know (somehow) that $t \leq \tilde{t}$.}
% Such a step can be important because this fact,
% which our method is in effect unaware of, may be essential to derive a desired bound; for example, if the input term is $\floor{t_1} - \floor{t_2}$ and we can derive $t_1 - t_2 \leq \tilde{t}$, then we are unable to derive the bound $0$, and 
% \twrchanged{
% instead are left with
% }
% the bound $\floor{t_1} - \floor{t_2} \leq t_1 - t_2 + 1 = 1$.
% In essence, this limitation is the cause of the discrepancy between our lower bound and the expert lower bound for elastic. \yf{does this make sense?}
% \twr{I find the existing explanation two paragraphs before the end of \sectref{overview} to be clearer.}
% \end{remark}
% \end{changebar}

% \begin{changebar}
The lower bound for NIRN that $\Tool$ computes does not seem desirable; it is complex, with an effective degree of 5, four floor terms, and three levels of floor nesting. This result may indicate that a saturation bound larger than $3$ is necessary, but we believe that the ultimate cause is inequational assumptions that are present in the original code but are missing in our model (this was our experience with the token example and the NIRN upper bound). 
%However, when we tried one candidate assumption, $\Tool$ did not terminate in over 16 hours.
% \twr{\delete{Say how long.}
% By the way, in general, we don't seem to be letting $\Tool$ run for very long.
% }
% \end{changebar}

\subsubsection{Sufficient Saturation Depth}
\label{Se:EvalQualitySatDepth}

The smallest saturation bound that suffices to produce the bounds of Table~\ref{Ta:EvalQuality} are marked on~\figref{SatScalability} using a larger marker of the same type is the other data points of the benchmark. All are below 3, and 2 suffices for some.

% \subsection{Modifying Product Saturation Depth}
% \begin{table}[t!]
% 	\centering
% 	{\small \caption{\label{Ta:ReqDepth}
%             {\small
% 	}}}
% \begin{tabular}{|| c | c ||}
% \hline
% Benchmark name & Required Depth\\
% \hline\hline
% elastic & 3/4\\
% fixed point & 3\\
% manual price & 2\\
% token & 3\\
% \hline
% \end{tabular}
% \end{table}

%% file: basic_table.tex
\begin{table}[t!]
	\centering
	{\small \caption{\label{Ta:Rewriting}
            {\small Performance of \Tool on the examples. 
            \#eq and \#ineq's are resp.\ equality and inequality assumptions initially given (not including instantiated axioms);
            \#floors is the number of integer divisions (floor of division) terms in the assumptions.
            \#c-eq and \#c-ineq are resp.\ the number of equalities/inequalities in the generated cone's ideal/polyhedron; \#m is the number of distinct monomials in the inequalities.
            time is the overall execution time of $\Tool$ (all times in seconds).
            csat is the time to saturate the cone. 
            reduce is the time to reduce w.r.t.\ the cone using local projection. 
            reduce-lp is the time to reduce using linear programming instead of local projection.
            All experiments in this table were taken using a product saturation depth of 3. 
	}}}
	\resizebox{.99\textwidth}{!}{
\begin{tabular}{|| l | r | r | r || r | r | r || r | r | r || r ||}
\hline
Benchmark name & \#eq & \#in & \#floors & \#c-eq & \#c-in & \#c-m & time (s) & csat (s) & reduce (s) & reduce-lp (s) \\
\hline\hline
elastic & 5 & 3 & 3 & 8 & 814 & 413 & 2.4 & 1.4 & 1.0 & 8.5\\
\hline

fixedPointInt & 0 & 2 & 2 & 2 & 162 & 131 & 0.3 & 0.2 & 0.1 & 0.7\\
\hline

manualPrice & 3 & 4 & 1 & 4 & 163 & 168 & 0.5 & 0.4 & 0.1 & 0.6\\
\hline

manualPriceMonotone & 6 & 5 & 2 & 8 & 218 & 228 & 0.7 & 0.6 & 0.1 & 1.7\\
\hline

tokent & 10 & 4 & 3 & 13 & 815 & 288 & 2.0 & 1.1 & 0.9 & 2.8\\
\hline

nirn & 10 & 5 & 6 & 16 & 4057 & 1351 & 81.0 & 9.6 & 71.4 & 963.1\\
\hline

\end{tabular}
}
\end{table} 

%% file: sat_size_table.tex
\begin{table}[t!]
	\centering
	{\small \caption{\label{Ta:SatScaleConeSize}
            {\small
            The size of the cone and run-time breakdown of each examples as the saturation depth is increased until a timeout of 10 minutes.
            \#c-eq and \#c-ineq are resp.\ the number of equalities/inequalities in the generated cone's ideal/polyhedron; \#m is the number of distinct monomials in the inequalities.
            time is the overall execution time of $\Tool$ (all times in seconds).
            csat is the time to saturate the cone. 
            reduce is the time to reduce w.r.t.\ the cone using local projection.
            -- indicates a timeout.
	}}}
\begin{tabular}{|| c | l || c | c | c | c | c | c | c |}\hline\multicolumn{2}{| c |}{\multirow{2}{*}{Benchmark}} & \multicolumn{7}{ | c | }{Saturation depth}
\\
\cline{3-9}\multicolumn{2}{| c |}{}  & 1 & 2 & 3 & 4 & 5 & 6 & 7\\
\hline\hline
\multirow{6}{*}{fixedPointInt} &\#c-eq & 2 & 2 & 2 & 2 & 2 & 2 & 2\\
\cline{2-9}
 & \#c-ineq & 8 & 42 & 162 & 489 & 1281 & 2993 & 6425\\
\cline{2-9}
 &  \#m & 9 & 41 & 131 & 336 & 742 & 1470 & 2682\\
\cline{2-9}
 &  time & 0.1 & 0.2 & 0.4 & 1.1 & 4.6 & 25.3 & 83.2\\
\cline{2-9}
 &  csat & 0.1 & 0.2 & 0.3 & 0.7 & 2.0 & 6.4 & 20.8\\
\cline{2-9}
 &  reduce & 0.0 & 0.0 & 0.1 & 0.4 & 2.5 & 18.9 & 62.4\\
\hline\hline
\multirow{6}{*}{manualPrice} &\#c-eq & 4 & 4 & 4 & 4 & 4 & 4 & 4\\
\cline{2-9}
 & \#c-ineq & 7 & 43 & 163 & 492 & 1284 & 2999 & 6431\\
\cline{2-9}
 &  \#m & 10 & 49 & 168 & 462 & 1092 & 2310 & 4488\\
\cline{2-9}
 &  time & 0.2 & 0.3 & 0.6 & 2.5 & 11.4 & 66.0 & 338.9\\
\cline{2-9}
 &  csat & 0.2 & 0.2 & 0.5 & 1.3 & 4.4 & 16.7 & 61.9\\
\cline{2-9}
 &  reduce & 0.0 & 0.0 & 0.1 & 1.2 & 7.0 & 49.3 & 277.0\\
\hline\hline
\multirow{6}{*}{manualPriceMonotone} &\#c-eq & 8 & 8 & 8 & 8 & 8 & 8 & --\\
\cline{2-9}
 & \#c-ineq & 8 & 53 & 218 & 712 & 1999 & 5001 & --\\
\cline{2-9}
 &  \#m & 11 & 60 & 228 & 690 & 1782 & 4092 & --\\
\cline{2-9}
 &  time & 0.2 & 0.4 & 0.8 & 4.3 & 47.9 & 283.1 & --\\
\cline{2-9}
 &  csat & 0.2 & 0.3 & 0.6 & 2.3 & 11.3 & 54.1 & --\\
\cline{2-9}
 &  reduce & 0.0 & 0.1 & 0.2 & 2.0 & 36.5 & 229.0 & --\\
\hline\hline
\multirow{6}{*}{tokent} &\#c-eq & 13 & 13 & 13 & -- & -- & -- & --\\
\cline{2-9}
 & \#c-ineq & 12 & 90 & 815 & -- & -- & -- & --\\
\cline{2-9}
 &  \#m & 12 & 71 & 288 & -- & -- & -- & --\\
\cline{2-9}
 &  time & 0.2 & 0.3 & 2.3 & -- & -- & -- & --\\
\cline{2-9}
 &  csat & 0.2 & 0.3 & 1.2 & -- & -- & -- & --\\
\cline{2-9}
 &  reduce & 0.0 & 0.1 & 1.0 & -- & -- & -- & --\\
\hline\hline
\multirow{6}{*}{nirn} &\#c-eq & 16 & 16 & 16 & -- & -- & -- & --\\
\cline{2-9}
 & \#c-ineq & 21 & 322 & 4057 & -- & -- & -- & --\\
\cline{2-9}
 &  \#m & 21 & 209 & 1351 & -- & -- & -- & --\\
\cline{2-9}
 &  time & 0.4 & 1.0 & 65.7 & -- & -- & -- & --\\
\cline{2-9}
 &  csat & 0.4 & 0.9 & 9.6 & -- & -- & -- & --\\
\cline{2-9}
 &  reduce & 0.0 & 0.1 & 56.1 & -- & -- & -- & --\\
\hline\hline
\end{tabular}
\end{table}

%% file: relatedwork.tex
\section{Related Work}\label{Se:related}

\paragraph{Optimization modulo theories}

Optimization modulo theories is the problem of minimizing (or maximizing) an objective function subject to a feasible region that is defined by a first-order formula \cite{ST:2012,LAKGC:2014,BCGIJRST:2021}.  This paper investigates a variation of this problem, in which the desired result is a \textit{term} rather than a value.  
Most work on optimization modulo theories is concerned with linear arithmetic.  An exception is \citet{BCGIJRST:2021},
 which---similarly to our work--handles non-linearity by ``linearizing'' the problem.  \citet{BCGIJRST:2021} incorporates a linear optimization modulo theories solver into an abstraction-refinement loop with an exact non-linear solver that generates lemmas on demand.  In contrast, our method eagerly communicates lemmas between a linear solver and a cooporating (incomplete) non-linear solver (based on \textit{cones of polynomials}).
 
\paragraph{Non-linear abstract domains}
The set of \textit{cones} can be seen as an abstract domain, in which elements are conjunctions of polynomial inequalities.  In this sense, cones are analogous to convex polyhedra, which represent conjunctions of linear inequalities.  Unlike the case of convex polyhedra, the concretization function that maps cones to their solutions is not injective; that is, ``inequivalent'' cones may represent the same set of points (e.g., the cones generated by $\{x,y\}$ and $\{x,xy\}$ are different, but the formulas $x \geq 0 \land y \geq 0$ and $x \geq 0 \land xy \geq 0$ represent the same points).   The saturation steps (\sectref{saturation}) can be conceived as semantic reduction \cite{CC:1979}: each saturation step derives new valid inequalities, while leaving the \textit{solutions} to those inequalities unchanged.

There are a number of non-linear abstract domains that include semantic-reduction steps akin to cone saturation \cite{KCBR:2018,BRZ:2005,GG:2008,CL:2005}.  The \textit{wedge} abstract domain introduced in \citet{KCBR:2018} is most similar to our saturation step---it uses a combination of Gr\"{o}bner-basis, congruence-closure, and polyhedral techniques for saturation.  Unlike the wedge domain, this paper uses a \textit{systematic} rather than \textit{heuristic} method for applying the rule that the product of two non-negative polynomials is non-negative.   In this regard, our method is similar to the domain of polynomial inequalities proposed in \citet{BRZ:2005}, and the domain of of polynomial equalities proposed in \cite{Colon:2004}.

\paragraph{Positivstellensatz}\label{Se:postivestull} In the case of real arithmetic positivstellensatz theorems \cite{putinar,Schmudgen} give completeness results concerning which polynomials are positive given a set of inequality assumptions. They are the non-linear analogue of Farkas' lemma. For Farkas' lemma conical combinations by non-negative scalars is sufficient; however positivstellensatz theorems consider an ideal and conical combinations by more general terms; e.g. sum-of-squares polynomials in the case of \citet{putinar}. In the Kriving--Stengle positivstellensatz, sum-of-squares polynomial combinations are taken with respect to products of the basis inequalities, which is similar to our process of taking products in~\sectref{takingProducts}. \citet{CHA:2020,FZJZX:2017} have both used positivstellensatz theorems with a degree bound in the context of program analysis by reducing to semi-definite programming, and consequently give numerical solutions. 
Our method does not perform such a reduction and can therefore easily give exact results. However, we lack an analogue of their completeness results.
%\paragraph{Rewriting}
%\cite{Kapur:1997,Kapur:2021}
%\zak{Need to invesitigate further -- \cite{Kapur:2021} may also solve the Gr\"{o}bner basis + uninterpreted functions problem in the same way.  (Shostak might also do this?)  Using purification + rewriting to compute congruence closure is due to \cite{Kapur:1997}}